\documentclass[twocolumn,english,prb]{revtex4-1}
\usepackage[T1]{fontenc}
\usepackage[latin9]{inputenc}
\usepackage{units}
\usepackage{amsmath}
\usepackage{graphicx}
\usepackage{amssymb}
\usepackage{esint}

\makeatletter

\providecommand{\tabularnewline}{\\}
\newcommand{\lyxdot}{.}

 \@ifundefined{textcolor}{}
 {%
   \definecolor{BLACK}{gray}{0}
   \definecolor{WHITE}{gray}{1}
   \definecolor{RED}{rgb}{1,0,0}
   \definecolor{GREEN}{rgb}{0,1,0}
   \definecolor{BLUE}{rgb}{0,0,1}
   \definecolor{CYAN}{cmyk}{1,0,0,0}
   \definecolor{MAGENTA}{cmyk}{0,1,0,0}
   \definecolor{YELLOW}{cmyk}{0,0,1,0}
 }

\makeatother

\usepackage{babel}

\makeatother

\usepackage{babel}

\begin{document}

\title{Doping dependence of femtosecond quasi-particle relaxation dynamics
in Ba(Fe,Co)$_{2}$As$_{2}$ single crystals: possible evidence for
normal state nematic fluctuations}

\author{L. Stojchevska$^{1}$, T. Mertelj$^{1}$, Jiun-Haw Chu$^{2,3}$,
Ian R. Fisher$^{2,3}$ and D. Mihailovic$^{1}$}

\affiliation{$^{1}$Complex Matter Dept., Jozef Stefan Institute, Jamova 39, Ljubljana,
SI-1000, Ljubljana, Slovenia }

\affiliation{$^{2}$Geballe Laboratory for Advanced Materials and Department of
Applied Physics, Stanford University, Stanford, California 94305,
USA}

\affiliation{$^{3}$Stanford Institute for Materials and Energy Sciences, SLAC
National Accelerator Laboratory, 2575 Sand Hill Road, Menlo Park,
California 94025, USA}

\date{\today}
\begin{abstract}
We systematically investigate the photoexcited (PE) quasi-particle
(QP) relaxation and low-energy electronic structure in electron doped
Ba(Fe$_{1-x}$Co$_{x}$)$_{2}$As$_{2}$ single crystals as a function
of Co doping, $0\leq x\leq0.11$. The evolution of the photoinduced
reflectivity transients with $x$ proceeds with no abrupt changes.
In the orthorhombic spin-density-wave (SDW) state a bottleneck associated
with a partial charge-gap opening is detected, similar to previous
results in different SDW iron-pnictides. The relative charge gap magnitude
$2\Delta(0)/k_{\mathrm{B}}T_{\mathrm{s}}$ decreases with increasing
$x$. In the superconducting (SC) state an additional relaxational
component appears due to a partial (or complete) destruction of the
SC state proceeding on a sub-0.5-picosecond timescale. From the SC
component saturation behavior the optical SC-state destruction energy,
$U_{\mathrm{p}}/k_{\mathrm{B}}=0.3$ K/Fe, is determined near the
optimal doping. The subsequent relatively slow recovery of the SC
state indicates clean SC gaps. The $T$-dependence of the transient
reflectivity amplitude in the normal state is consistent with the
presence of a pseudogap in the QP density of states. The polarization
anisotropy of the transients suggests that the pseudogap-like behavior
might be associated with a broken point symmetry resulting from nematic
electronic fluctuations persisting up to $T\simeq200$ K at any $x$.
The second moment of the Eliashberg function, obtained from the relaxation
rate in the metallic state at higher temperatures, indicates a moderate
electron phonon coupling, $\lambda\lesssim0.3$, that decreases with
increasing doping.
\end{abstract}
\maketitle

\section{Introduction}

Very soon after the discovery of high-temperature superconductivity
in iron-based pnictides \cite{KamiharaKamihara2006,kamiharaWatanabe2008,RenChe2008}
some unusual properties of the normal state were indicated by various
experimental techniques\cite{LiuWu2008,AhilanNing2009,NingAhilan2009,HessKondrat2009,MerteljKabanov2009prl,ChuAnalytis2010,ChuangAllan2010,TanatarThaler2010,DuszaLucarelli2011}
indicating a possible pseudogap.\cite{AhilanNing2009,MerteljKabanov2009prl,NingAhilan2009,ChuangAllan2010}
More recently, the presence of the normal-state electronic nematic
fluctuations\cite{ChuAnalytis2010,ChuangAllan2010,TanatarBlomberg2010,DuszaLucarelli2011,YingWang2011,YiLu2011}
has been suggested from a remarkable anisotropy of physical properties
induced by application of an external uniaxial stress in the tetragonal
phase, well above the structural phase transition. 

In comparison to the cuprates, where the pseudogap is ubiquitous,
the existence of a pseudogap in iron-based pnictides is still controversial
since no strong anomalies are present in the normal-state in-plane
transport properties.\cite{Rullier-AlbenqueColson2009,LeeBartkowiak2009}
As in the cuprates, it is believed, that understanding the unusual
normal state might be a key for revealing the origin of high temperature
superconductivity observed in these systems.

Time domain optical spectroscopy has been, among other spectroscopies,
very instrumental in elucidating the nature of the unusual normal
state in the cuprates by virtue of the fact that different components
in the low-energy excitation spectrum could be distinguished by their
different lifetimes.\cite{DemsarPodobnik1999,KaindlWoerner2000,AverittRodriguez2001,SegreGedik2002,KusarDemsar2005,LiuToda2008,CaoWei2008}
In iron pnictides several reports on photoexcited carrier dynamics
exist\cite{MerteljKabanov2009prl,MerteljKabanov2009jsnm,MerteljKusar2010,TorchinskyChen2010,ChiaTalbayev2010,StojchevskaKusar2010,GongLai2010,MansartBoschetto2010,TorchinskyMcIver2011},
but   doping dependence studies are still incomplete\cite{MerteljKusar2010,ChiaTalbayev2010,TorchinskyMcIver2011}.

Here we present a systematic temperature ($T$) dependent time-domain
optical spectroscopy study in Ba(Fe$_{1-x}$Co$_{x}$)$_{2}$As$_{2}$
spanning a large part of the $x-T$ phase diagram from the undoped
spin density wave (SDW) metallic state at $x=0$ through coexisting
superconducting-SDW state around $x=5\%$, to the overdoped superconducting
(SC) state at $x=11\%$. We investigate the photoexcited quasiparticle
relaxation in the low-$T$ SDW and SC ground states as well as in
the intermediate-$T$ {}``pseudogap'' state and the room-$T$ metallic
state. 

From the temperature dependencies of the optical relaxation transients
we infer the relaxation bottlenecks which we attribute to opening
of the charge gap with the BCS-like temperature dependence in the
SDW state and the presence of a pseudogap-like suppression of the
electronic density of states at higher temperatures. Surprisingly,
we find that the 2-fold optical symmetry observed in the orthorhombic
SDW state persists well into the tetragonal state, suggesting association
of the pseudogap-like behavior with a broken point symmetry and the
presence of nematic fluctuations. 

The analysis of the high temperature relaxation dynamics experimentally
confirms the supposition of a moderate electron phonon coupling in
iron-pnictide superconductors.

We discuss separately the response of the SC state, where we observe
an ultrafast nonthermal destruction of the SC condensate. The subsequent
SC state recovery dynamics indicates clean gaps with the BCS-like
temperature dependence.

The paper starts with presentation and description of experimental
data in Section II, followed by a more detailed analysis and modeling,
separately focusing on different regions of the phase diagram, in
Section III. Conclusions and summary are presented in Section IV.

\begin{figure}[tbh]
\begin{centering}
\includegraphics[width=0.3\textwidth,angle=-90]{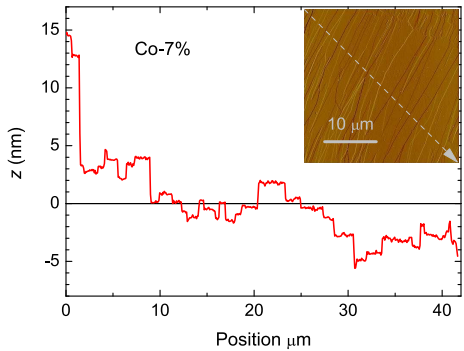} 
\par\end{centering}

\caption{Atomic force microscope analysis of the cleaved surface quality in
the near optimally doped Co-7\% sample. The dashed arrow indicates
the location of the plotted $z$ profile.}

\label{fig:fig-surface} 
\end{figure}
\begin{figure*}
\includegraphics[scale=0.6,angle=-90]{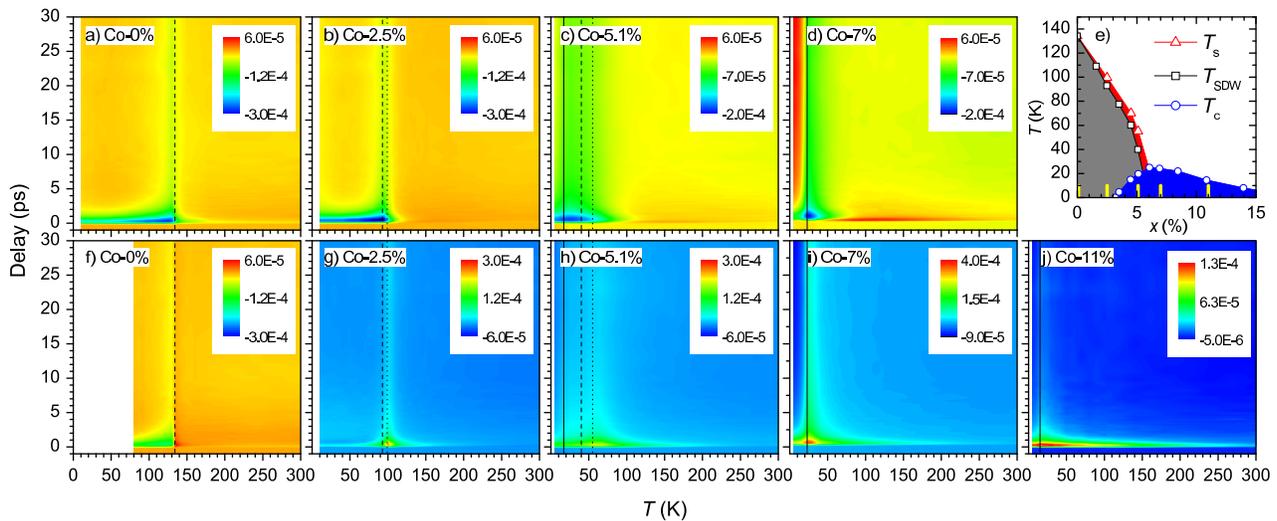}

\caption{(Color online) Density plots of $\Delta R/R$ transients at 13 $\mu$J/cm$^{2}$
pump fluence as a function of temperature at different dopings and
probe polarizations (a)-(d), (f)-(j). The top row (a)-(d) corresponds
to the $\mathcal{P}^{-}$ and the bottom row (f)-(i) to the $\mathcal{P}^{+}$
polarization. The vertical lines indicate $T\mathrm{_{c}}$ (full
lines), $T\mathrm{_{SDW}}$ (dashed lines) and $T\mathrm{_{s}}$ (dotted
lines). The phase diagram\cite{ChuAnalytis2009,LesterChu2009} of
Ba(Fe$_{1-x}$Co$_{x}$)$_{2}$As$_{2}$ (e). The investigated-samples
dopings are marked by the yellow vertical bars.}
\label{fig:DR-2D}
\end{figure*}

\section{Experimental}

\subsection{Setup and samples}

Single crystals of Ba(Fe$_{1-x}$Co$_{x}$)$_{2}$As$_{2}$ with Co
dopings, $x=0$\%, 2.5\%, 5.1\%, 7\% and 11\%, were grown from a self
flux, and characterized as described previously.\cite{ChuAnalytis2009,ChuAnalytis2010,DuszaLucarelli2011}
The Co-0\% and Co-2.5\% samples are underdoped and non-superconducting
with SDW ordering while the Co-5.1\% sample exhibits a coexistence
of SDW and superconductivity at low $T$ {[}see Fig. \ref{fig:DR-2D}
(e){]}. The Co-7\% and Co-11\% samples correspond to the near optimally
doped and overdoped region of the SC phase diagram with no SDW ordering. 

For optical measurements the crystals were glued onto a copper sample
holder and cleaved by a razor before mounting in an optical liquid-He
flow cryostat. Cleaving resulted in a terrace like surface (see Fig.
\ref{fig:fig-surface}) with a typical terrace width of a few $\mu$m
and step height of a few nm. The relative orientation of the terraces
with respect to the crystal axes was not determined.

Measurements of the photoinduced reflectivity, $\Delta R/R$, were
performed using the standard pump-probe technique, with 50 fs optical
pulses from a 250-kHz Ti:Al$_{2}$O$_{3}$ regenerative amplifier
seeded with an Ti:Al$_{2}$O$_{3}$ oscillator. Unless otherwise noted,
we used the pump photons with the doubled ($\hbar\omega_{\mathrm{P}}=3.1$
eV) photon energy and the probe photons with the laser fundamental
1.55 eV photon energy to easily suppress the scattered pump photons
by long-pass filtering. In some cases we used (for comparison) also
the degenerate pump-photon energy of $1.55$ eV. The pump and probe
beams were nearly perpendicular to the cleaved sample surface with
polarizations perpendicular to each other%
\footnote{We observed no pump polarization dependence of the response at fixed
probe polarization.%
} and oriented with respect to the the crystals to obtain the maximum/minimum
amplitude of the response at low temperatures. The pump and probe
beam diameters were determined by measuring the transmittance of calibrated
pinholes mounted at the sample place\cite{KusarKabanov2008} resulting
in 60$\mu$m/50$\mu$m and 70$\mu$m/40$\mu$m for 3.1eV/1.55eV and
1.55eV/1.55eV pump/probe energies, respectively.

\subsection{Overview of the experimental data set}

\begin{figure}[tbh]
\begin{centering}
\includegraphics[width=0.48\textwidth]{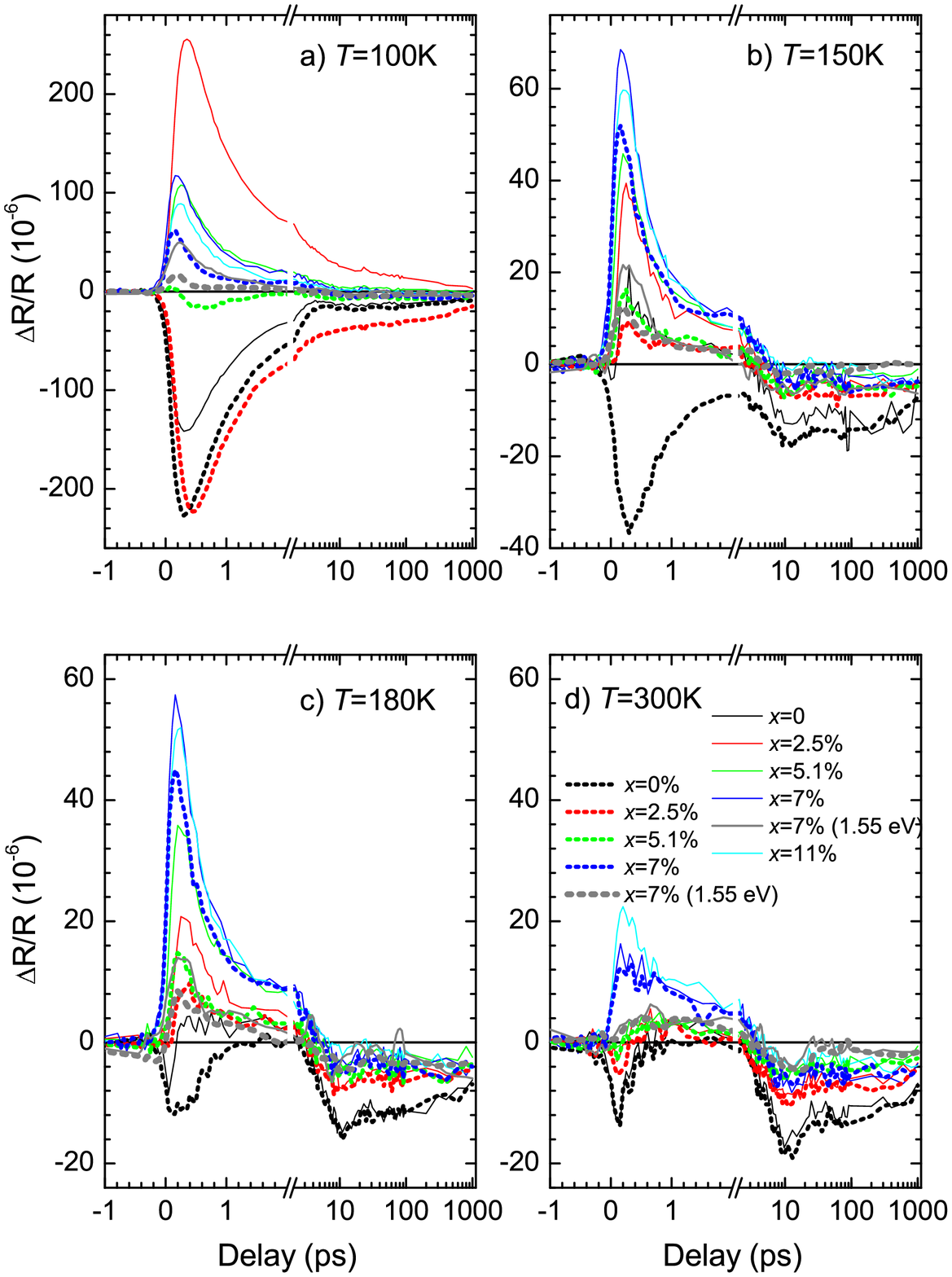} 
\par\end{centering}

\caption{(Color online) Anisotropy of the raw $\Delta R/R$ transients at different
temperatures as a function of Co doping $x$ at 13 $\mu$J/cm$^{2}$
pump fluence. The full and dotted lines correspond to $\mathcal{P}^{+}$
and $\mathcal{P}^{-}$ polarizations, respectively. The gray lines
are the transients measured with 1.55 eV pump photon energy at 11
$\mu$J/cm$^{2}$ in the Co-7\% sample.}

\label{fig:fig-DRvsX} 
\end{figure}

In Fig. \ref{fig:DR-2D} we plot the temperature dependence of the
raw photoinduced reflectivity ($\Delta R/R$) transients at different
dopings. Despite no deliberate uniaxial strain was applied to the
samples\cite{ChuAnalytis2010} all samples except the Co-11\% sample
showed a 2-fold anisotropy with respect to the probe polarization
below $\sim$200K (see Fig. \ref{fig:fig-DRvsX}). Since, according
to the optical penetration depth in iron-pnictides,\cite{MerteljKusar2010}
the probed volume is limited to a few tens of nanometers thick layer
near the surface we attribute the anisotropy to an anisotropic surface-strain
bias induced by the local pump-beam thermal load in the presence of
the uni-directionally ordered terraces shown in Fig. \ref{fig:fig-surface}.
In the absence of information which crystallographic direction correspond
to the two different orthogonal probe polarization we denote the polarization
corresponding to the low-temperature minimal and maximal sub-picosecond
peak $\Delta R/R$ value $\mathcal{P}^{-}$ and $\mathcal{P}^{+}$,
respectively.

\begin{figure}[tbh]
\begin{centering}
\includegraphics[width=0.48\textwidth]{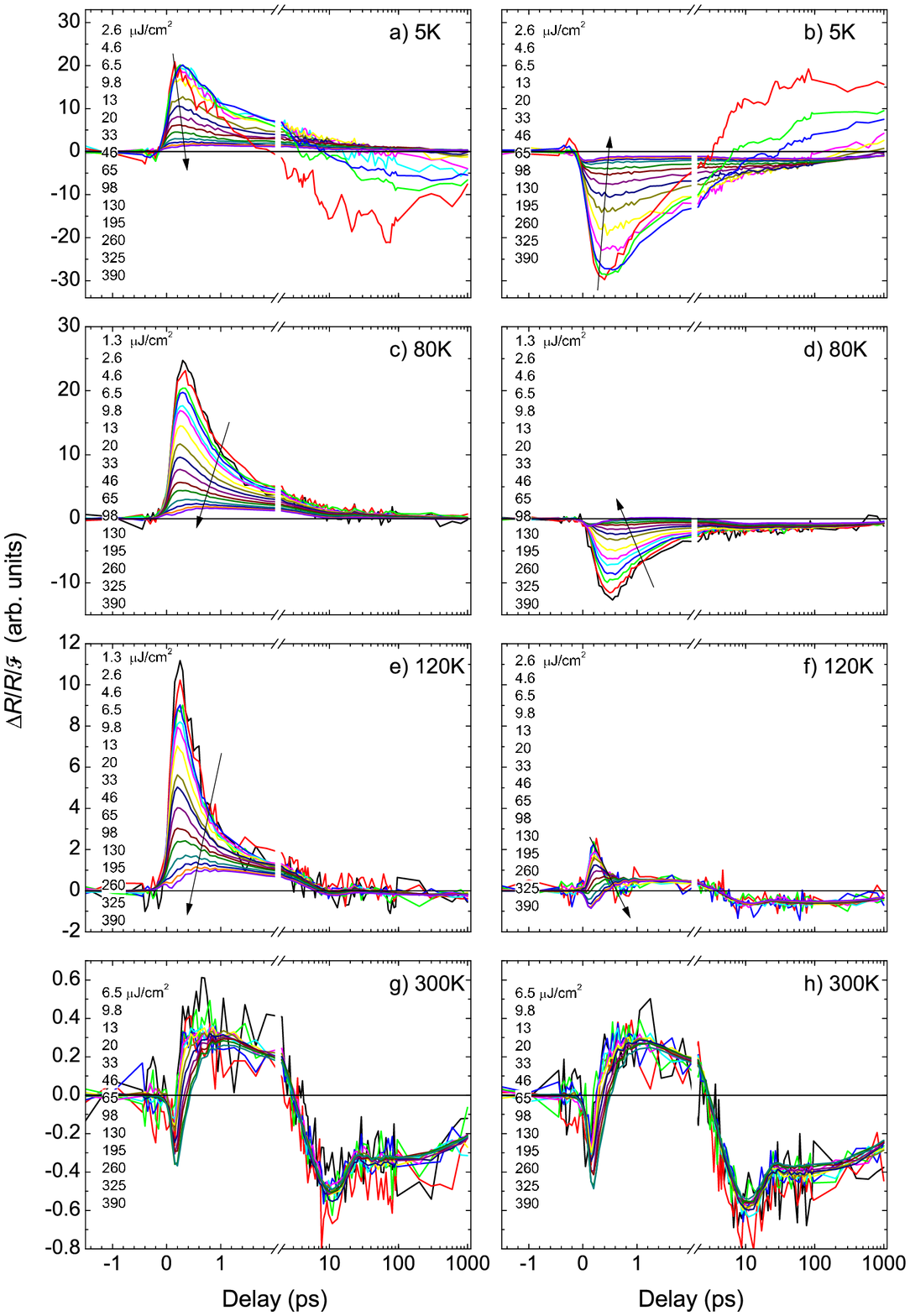} 
\par\end{centering}

\caption{(Color online) Fluence-normalized $\Delta R/R/\mathcal{F}$ transients
as a function of $\mathcal{F}$ at different temperatures and probe
polarizations in the Co-5.1\% sample. The arrows indicate the direction
of increasing $\mathcal{F}$. Overlapping curves indicate a linear
$\mathcal{F}$ dependence. The left and right columns correspond to
the $\mathcal{P}^{+}$ and $\mathcal{P}^{-}$ polarization, respectively.}

\label{fig:fig-DR-vs-F-5.1=000025} 
\end{figure}

At any doping the $\Delta R/R$ transients show a saturation with
increasing pump laser fluence ($\mathcal{F}$) at low temperatures.
At high temperatures the saturation behavior vanishes as shown as
an example for the Co-5.1\% sample in Fig. \ref{fig:fig-DR-vs-F-5.1=000025}. 

\subsection{Results in samples that show SDW ordering}

The response in the undoped Co-0\% sample is very similar to previous
results in undoped SrFe$_{2}$As$_{2}$ and SmAsFeO.\cite{MerteljKusar2010,StojchevskaKusar2010}
Below $T_{\mathrm{s}}=T_{\mathrm{SDW}}$ the transients are dominated
by the initial single exponential relaxation (see Fig. \ref{fig:fig-DRvsX}).
At $T_{\mathrm{s}}$ a slowing down of relaxation is observed in the
form of a long lived relaxation, which is following the initial $\sim1.5$
ps exponential decay and extends throughout the experimental ns time
window {[}see $x=2.5\%$ curves in Fig. \ref{fig:fig-DRvsX} (a){]}.
Above $T_{\mathrm{s}}$ the amplitude of the initial sub-ps relaxation
strongly drops and the structure at around 10 ps, which was observed
also in previously studied iron-pnictides,\cite{StojchevskaKusar2010,MerteljKusar2010}
becomes apparent. The structure could be associated with the acoustic
wave propagating into the sample after expansion of the excited volume
due to the transient laser-pulse heating and will not be discussed
further.

\begin{figure}[tbh]
\begin{centering}
\includegraphics[width=0.45\textwidth]{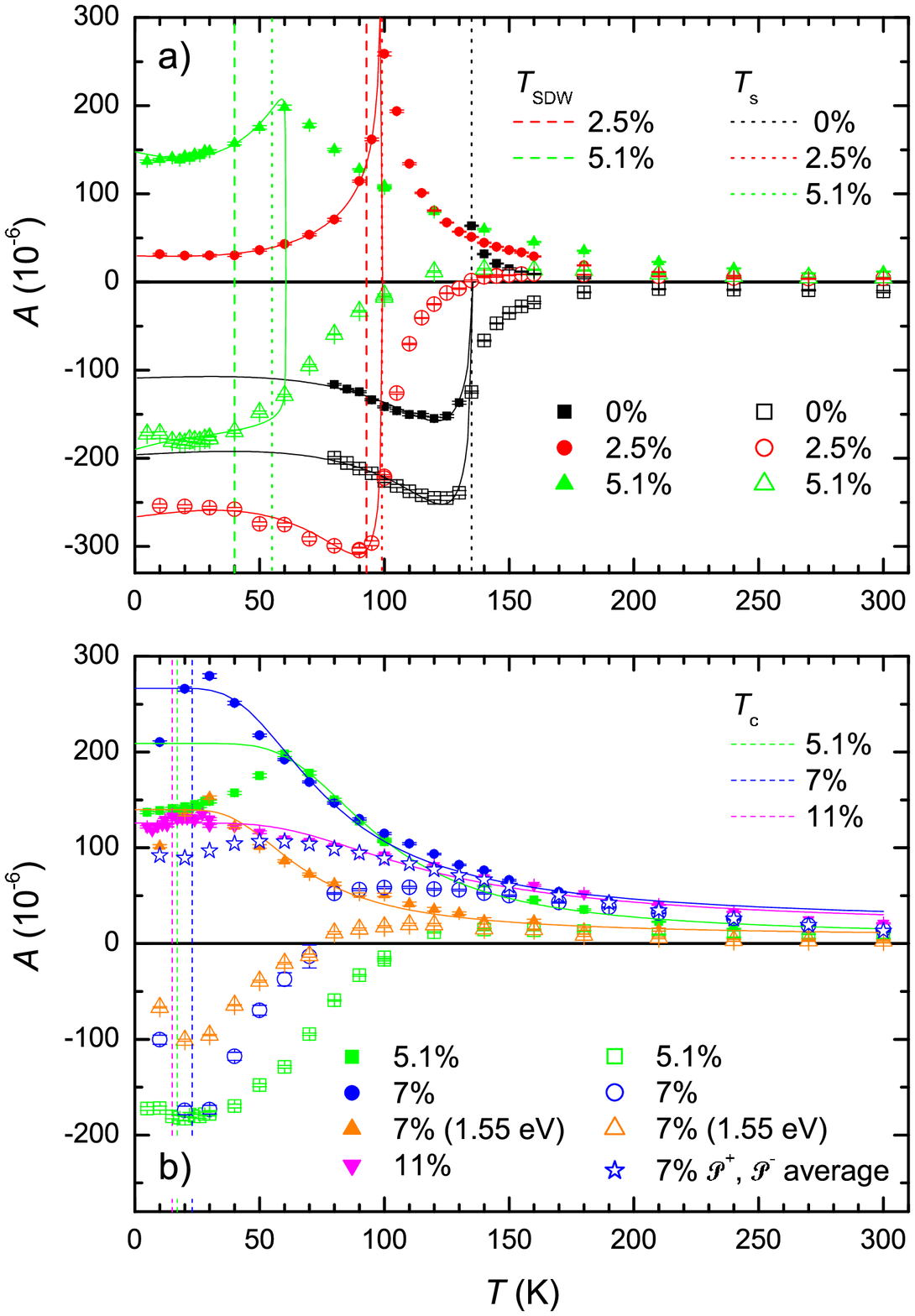} 
\par\end{centering}

\caption{(Color online) Amplitudes of the raw reflectivity transients as a
function of temperature at different Co dopings, $x$, at 13 $\mu$J/cm$^{2}$
pump fluence. The full and open symbols correspond to $\mathcal{P}^{+}$
and $\mathcal{P}^{-}$ polarizations, respectively. Open stars represent
the amplitude of the probe-polarization averaged transients in the
Co-7\% sample. The thin lines in (a) are fits of equation (\ref{eq:DR-npe})
discussed in text. The thin lines in (b) are the $T$-independent
gap fits (\ref{eq:AvsT-PG}) discussed in text. The vertical lines
indicate $T\mathrm{_{c}}$ (obtained from our optical data), $T\mathrm{_{SDW}}$
and $T\mathrm{_{s}}$ (obtained from the phase diagram\cite{ChuAnalytis2009,LesterChu2009}).}

\label{fig:fig-AvsT-2-All} 
\end{figure}

With increasing Co doping the anomalies around $T_{\mathrm{s}}$ become
broader {[}see Fig. \ref{fig:fig-AvsT-2-All} (a){]} and the drop
of the amplitudes with increasing $T$ above $T_{\mathrm{s}}$ becomes
slower. There are no clear separate features observed at the spin-density-wave
(SDW) transition temperature ($T_{\mathrm{SDW}}$) in the Co-2.5\%
and Co-5.1 samples, where the SDW transition is split from the structural
phase transition.

There is also a marked difference in the probe polarization anisotropy%
\footnote{The anisotropy indicates a preferential ordering of the orthorhombic
twin domains in the probed volume due to the anisotropic surface strain.%
} of the Co-0\% sample with respect to the Co-2.5\% and Co-5.1\% samples
where the amplitude shows either a peak for the $\mathcal{P}^{+}$
polarization or a step-like increase for the $\mathcal{P}^{-}$ polarization
at $T_{\mathsf{s}}$ {[}see Fig. \ref{fig:fig-AvsT-2-All} (a){]}.
This difference could be explained by a lower degree of detwinning
due to the surface-strain bias in the Co-0\% sample.

\subsection{Results in superconducting samples}

\begin{figure}[tbh]
\begin{centering}
\includegraphics[angle=-90,width=0.48\textwidth]{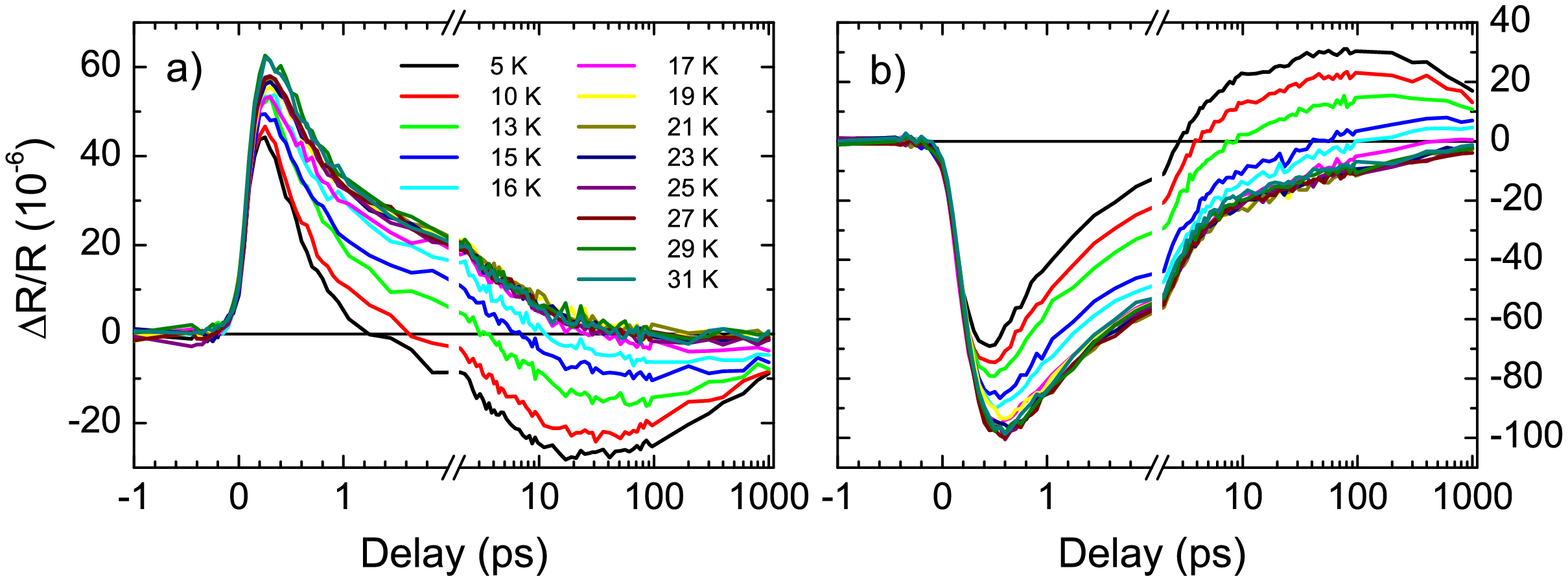} 
\par\end{centering}

\caption{(Color online) Low temperature $\Delta R/R$ transients as a function
of temperature in the Co-5.1\% sample at 3.9 $\mu$J/cm$^{2}$ pump
fluence for two orthogonal probe polarizations $\mathcal{P}^{+}$
(a) and $\mathcal{P}^{-}$ (b). }

\label{fig:fig-sc-5.1=000025} 
\end{figure}

In the superconducting samples ($x>2.5\%)$ an additional SC component
relaxing on a hundreds-of-picosecond timescale appears below the critical
temperature ($T\mathrm{_{c}}$) {[}see Fig. \ref{fig:fig-sc-5.1=000025}
(a) and (b){]}. The component is the most clearly observed at low
pump fluences and has the largest magnitude in the optimally doped
Co-7\% sample {[}see Fig. \ref{fig:DR-2D} (d) and (i){]}. 

As previously observed in the cuprates\cite{KusarKabanov2008} and
iron-pnictides\cite{MerteljKabanov2009prl,MerteljKusar2010} the SC
component saturates at a lower pump fluence than the components which
are present also above $T_{\mathrm{c}}$. The saturation of the SC
component is associated\cite{MerteljKabanov2009prl,MerteljKusar2010}
with a complete SC state destruction in the optically probed volume. 

In the normal state the transients in the near-optimally-doped Co-7\%
sample show a similar temperature evolution as in the Co-5.1\% sample
above $T_{\mathrm{s}}$, but shifted to lower temperatures. When cooling
from the room temperature at $\sim70K$ ($\sim100K$ in the Co-5.1\%
sample) the transients for the $\mathcal{P}^{-}$ polarization show
an emergence of a negative picosecond component resulting in a change
of sign together with appearance of the long lived relaxation tail
below $\sim50$ K ($\sim70$ K in the Co-5.1\% sample). There is,
however, no structural transition with a peak of the amplitude as
in the Co-5.1\% sample at $T_{\mathrm{s}}\simeq60$ K, but a direct
transition to the SC state at $T_{\mathrm{c}}\simeq23$ K%
\footnote{The critical temperature was determined from our optical measurements
based on the sample holder temperature and is apparently lower than
the phase diagram {[}Fig. \ref{fig:DR-2D} (e){]} value due to the
sample heating by the laser.%
}.

The transients in the overdoped Co-11\% sample show, on the other
hand, just a monotonous increase of the magnitude and relaxation timescale
when the temperature is lowered from the room temperature down to
the SC transition temperature similar to the probe polarization averaged
transients in the Co-7\% sample {[}see Fig. \ref{fig:fig-AvsT-2-All}
(b){]}.

\section{Analysis and Discussion}

\begin{figure}[tbh]
\begin{centering}
\includegraphics[width=0.33\textwidth,angle=-90]{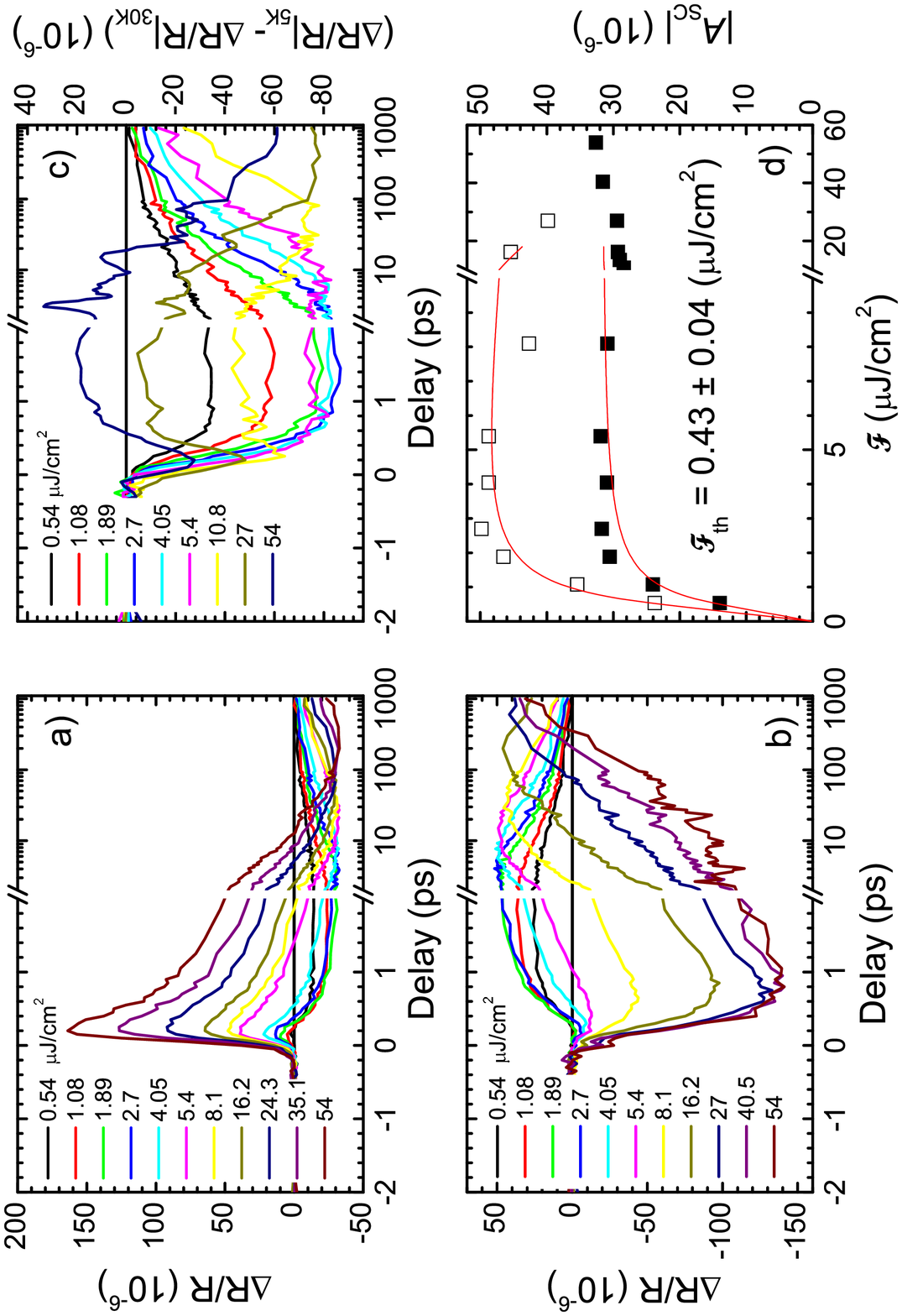} 
\par\end{centering}

\caption{The raw $\Delta R/R$ transients in the optimally doped Co-7\% sample
as a function of $\mathcal{F}$ at $\mathcal{P}^{+}$ (a) and $\mathcal{P}^{-}$
(b) polarizations. The SC component as a function of $\mathcal{F}$
for the $\mathcal{P}^{+}$ polarization (c). The $\mathcal{F}$-dependence
of the SC component amplitude for $\mathcal{P}^{+}$ (full symbols)
and $\mathcal{P}^{-}$ (open symbols) polarizations in the optimally
doped Co-7\% sample (d). The thin lines in (d) are fits of the non-homogeneous
saturation model.\cite{KusarKabanov2008} All data in this figure
were obtained at $T=5$K and 1.55 eV pump-photon energy.}

\label{fig:fig-Co-7=000025-F-dep} 
\end{figure}
\begin{figure}[tbh]
\begin{centering}
\includegraphics[width=0.48\textwidth]{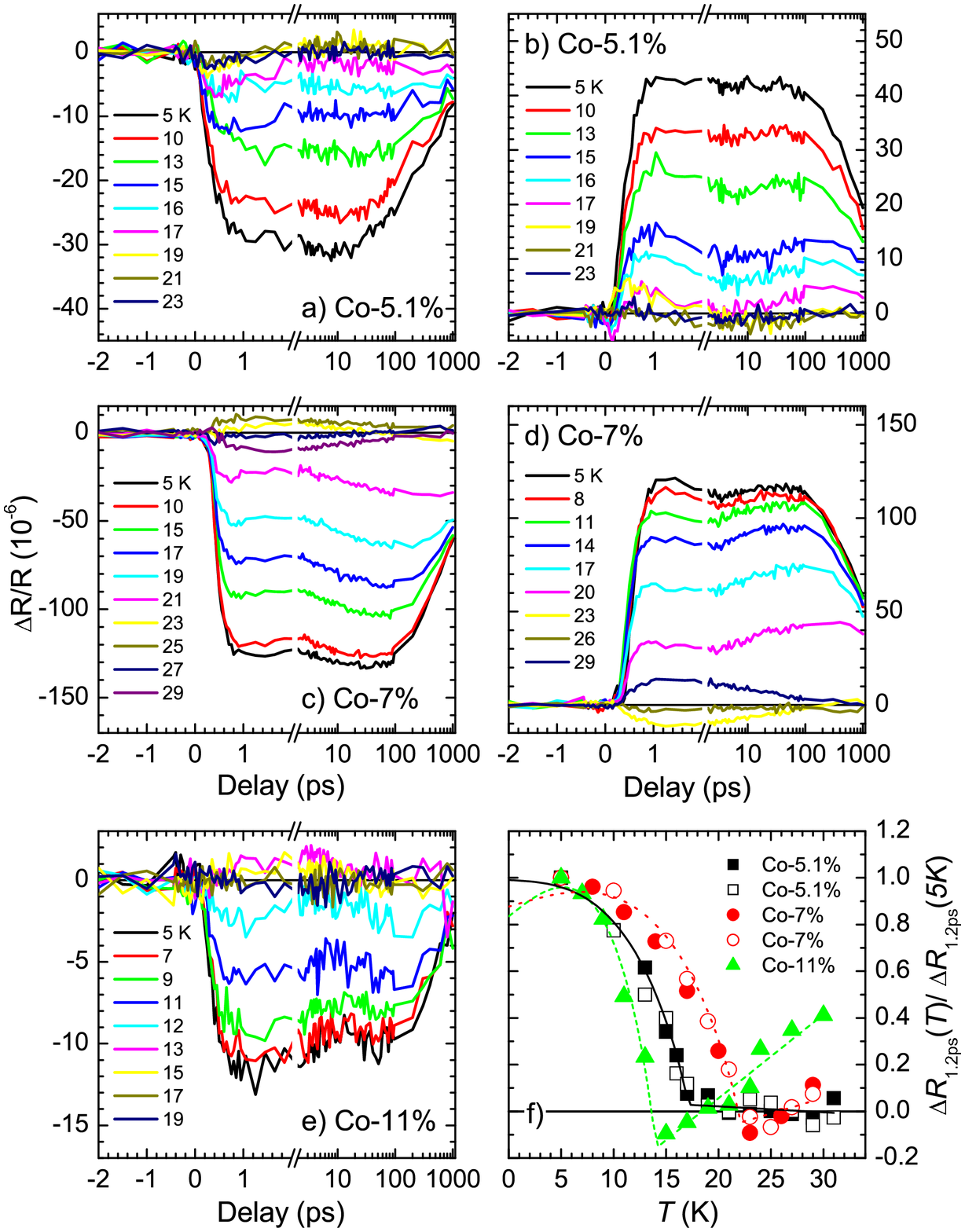} 
\par\end{centering}

\caption{(Color online) The superconducting response at different polarizations
in the Co-5.1\% sample at 3.9 $\mu$J/cm$^{2}$ pump fluence (a) and
(b), in the Co-7\% sample at 13 $\mu$J/cm$^{2}$ pump fluence (c)
and (d) and in the Co-11\% sample at 2.6 $\mu$J/cm$^{2}$ pump fluence
(e). The normalized photoinduced reflectivity at 1.2 ps in all superconducting
samples (f). Full and open symbols correspond to $\mathcal{P}^{+}$
and $\mathcal{P}^{-}$ probe polarizations, respectively. The lines
in (d) are Mattis-Bardeen fits (\ref{eq:MattBard}) discussed in text.}

\label{fig:fig-sc-all} 
\end{figure}

\subsection{SC state destruction and recovery}

\subsubsection{Fluence dependence}

Due to a finite noise magnitude we were able to investigate the $\mathcal{F}$-dependence
of the SC component below the saturation fluence only in the nearly
optimally doped sample ($x=7$\%) with the highest $T_{\mathrm{c}}\simeq23$
K. At the lowest $\mathcal{F}$ the SC component dominates the raw
$\Delta R/R$ transients {[}Fig. \ref{fig:fig-Co-7=000025-F-dep}
(a) and (b){]}. With increasing $\mathcal{F}$, however, the SC component
quickly saturates while an increasing contribution of the normal state
components starts to prevail at shorter timescales resulting in a
shift of the minimum corresponding to the saturated SC component for
the $\mathcal{P}^{-}$ probe polarization {[}Fig. \ref{fig:fig-Co-7=000025-F-dep}
(a){]} (and the maximum for the $\mathcal{P}^{+}$ probe polarization
polarization {[}Fig. \ref{fig:fig-Co-7=000025-F-dep} (b){]}) towards
longer delays. 

To extract the SC relaxation component we used the observation that
the $\Delta R/R$ transients only weakly depend on the temperature
just above $T_{\mathrm{c}}$.\cite{MerteljKabanov2009prl,MerteljKusar2010}
In Fig. \ref{fig:fig-Co-7=000025-F-dep} (c) we show the $\mathcal{F}$-dependence
of the SC component for the $\mathcal{P}^{+}$ probe polarization
obtained by subtraction of the normal state $\Delta R/R$ transients
measured above $T_{\mathrm{c}}$ (at 30K) from the transients measured
at 5K. The subtraction procedure clearly fails at high $\mathcal{F}$
producing an apparent non-monotonous temporal dependence of the SC
component at longer delays. The failure is attributed to the weak
$T$-dependence of the normal state components and/or a systematic
error, which become large in comparison to the magnitude of the saturated
SC component at higher $\mathcal{F}$. Nevertheless, one can observe
an increasing duration of the plateau corresponding to the transient
destruction of the SC state and slowing down of the subsequent SC
state recovery with increasing $\mathcal{F}$. 

We use the inhomogeneous SC-state destruction model\cite{KusarKabanov2008}
to fit $\mathcal{F}$-dependence of the SC component amplitudes {[}see
Fig. \ref{fig:fig-Co-7=000025-F-dep}(d){]} and determine the SC state
destruction threshold external fluence $\mathcal{F}_{\mathrm{T}}=0.43\pm0.04$
$\mu$J/cm$^{2}$. Taking optical constants from Ref. {[}\onlinecite{BarisicWu2010}{]}
we obtain the reflectivity $R$=0.37 and the optical penetration depth
$\lambda_{\mathrm{op}}=34$ nm at the 1.55-eV pump photon energy,
giving the bulk SC state destruction energy density required to completely
destroy the superconducting state: $\nicefrac{U_{\mathrm{p,Co-122}}}{k_{\mathrm{B}}}=\nicefrac{\mathcal{F}_{\mathrm{T}}\left(1-R\right)}{\lambda_{\mathrm{op}}k_{\mathrm{B}}}=0.3$
K/Fe ($U_{\mathrm{p,Co-122}}=4.9$ J/mole). This value is much smaller
than the energy necessary to heat the sample thermally to $T_{\mathrm{c}}$,
$\nicefrac{U_{\mathrm{Q}}}{k_{\mathrm{B}}}=\intop_{5K}^{T_{\mathrm{c}}}\nicefrac{c_{p}}{k_{\mathrm{B}}}dT\simeq2.4$
K/Fe ($U_{\mathrm{Q}}=40$ J/mole) indicating that the SC destruction
is highly non-thermal. On the other hand, the thermodynamic condensation
energy, $\nicefrac{U\mathrm{_{c,Co-122}}}{k_{\mathrm{B}}}=0.15\pm0.02$
K/Fe%
\footnote{We calculated $U_{\mathrm{c}}$ from the heat capacity data in Ref.
\onlinecite{HardyBurger2010}. %
} is only half of $U_{\mathrm{p,Co-122}}$ indicating that a half of
the optical energy, initially completely absorbed by the electronic
subsystem, is quickly (within $\mbox{\ensuremath{\tau}}_{\mathrm{r}}\simeq0.5$
ps) transferred to the sub-gap phonons, with $\hbar\omega_{\mathrm{ph}}<2\Delta_{\mathrm{SC}}$,
which can not break Cooper pairs.\cite{StojchevskaKusar2011} 

Comparing the destruction energy density with the near optimally doped
SmAsFe(O,F) ($T_{c}\simeq49K$).\cite{MerteljKusar2010} we find that
it is much smaller than $\nicefrac{U_{\mathrm{p,Sm-1111}}}{k_{\mathrm{B}}}=1.8$
K/Fe. The ratio of the destruction energies $\nicefrac{U_{\mathrm{p},Sm-1111}}{U_{\mathrm{p},Co-122}}=6$
is, however, close to the ratio of the critical temperatures squared,
$(\nicefrac{T_{\mathrm{c,Sm-1111}}}{T_{\mathrm{c,Co-122}}})^{2}=4.5$,
which corresponds to the ratio of the condensation energies if we
assume a similar SC gap structure and the electronic density of states
in both compounds.

\subsubsection{Temperature dependence}

To study the temperature dependence of the SC component we subtracted
the average of the normal-state $\Delta R/R$ transients up to $\sim$10K
above $T{}_{\mathrm{c}}$ from the raw transients. The resulting SC
component is shown in Fig. \ref{fig:fig-sc-all} (a-e) for different
Co dopings and polarizations. Due to a rather small signal to noise
ratio the $\mathcal{F}$-linear pump fluence region was not accessible
so $\mathcal{F}$ was chosen significantly above the SC component
saturation threshold in all cases. 

The SC component shows a rise-time of $\mbox{\ensuremath{\tau}}_{\mathrm{r}}\lesssim$
0.5 ps followed by a plateau extending from tens of picoseconds at
5K to several hundred picoseconds when the temperature is increased
towards $T_{\mathrm{c}}$.%
\footnote{The weak non monotonic temporal dependence of the SC component during
the plateau could not be reliably identified as an intrinsic effect
and is attributed to a weak $T$-dependence of the subtracted normal
state response.%
} As discussed above, the plateau corresponds to the transient destruction
of the SC state. The timescale of the SC state recovery following
the plateau is $\sim1$ ns at 5K and increases with increasing temperature. 

Except in the overdoped Co-11\% sample, which shows no polarization
dependence of the $\Delta R/R$ transients and the smallest magnitude
of the saturated SC component, the sign of the SC component changes
for the two orthogonal probe polarizations. There is, however, no
difference (within the experimental error) in the delay evolution
of the SC component among the $\mathcal{P}^{+}$ and $\mathcal{P}^{-}$
polarizations. Taking into account, that different polarizations probe
different parts of the Fermi surface,\cite{YiLu2011} this indicates
that the destruction and the recovery of the SC order parameter is
uniform along different parts of the Fermi surface.

In Fig. \ref{fig:fig-sc-all} (f) we plot the temperature dependence
of the SC-component saturated amplitude for all SC samples. The linear
$T$-dependence of the amplitude above $T_{\mathrm{c}}$ indicates
some residual contribution due to the weak $T$-dependence of the
non-SC contributions. Below $T_{\mathrm{c}}$ we observe (on top of
the linear $T$-dependence) the characteristic Mattis-Bardeen $T$-dependence
given by the high-frequency limit of the Mattis-Bardeen formula,\cite{MattisBardeen1985,MerteljKusar2010}
\begin{equation}
\frac{\Delta R}{R}\propto\left(\frac{\Delta\left(T\right)}{\hbar\omega}\right)^{2}\log\left(\frac{3.3\hbar\omega}{\Delta\left(T\right)}\right),\label{eq:MattBard}\end{equation}
where $\hbar\omega$ is the probe-photon energy and $\Delta\left(T\right)$
the superconducting gap with the BCS temperature dependence.

\subsubsection{Comparison with SmFeAs(O,F)}

The presence of the plateau and the slow SC state recovery is different
than in the near-optimally doped SmFeAs(O,F), where a two stage SC
recovery was observed.\cite{MerteljKusar2010} The fast equilibration
stage, appearing on a $\sim5$ ps timescale in SmFeAs(O,F), corresponds
to the initial local equilibration among all degrees of freedom and
the slow, appearing on a several-hundred-picosecond timescale, corresponds
to the energy escape from the optically probed volume by the diffusive
heat transport.\cite{MerteljKusar2010} The absence of the two-stage
relaxation in Ba(Fe$_{1-x}$Co$_{x}$)$_{2}$As$_{2}$ is consistent
with the fact, that, at the pump fluences used for the $T$-scans,
the total laser energy deposited in the optically probed volume corresponds
to heating the sample to a temperature well above $T_{\mathrm{c}}$,
in the 30-40K range. The sample thus remains in the normal state after
the fast stage and the SC recovery is governed by the diffusive-heat-transport
slow stage only. 

Moreover, in the present case a separate two-stage relaxation is not
observed even at the lowest $\mathcal{F}$ just above the threshold
{[}see Fig. \ref{fig:fig-Co-7=000025-F-dep} (c){]} suggesting that
the initial local equilibration is slower than in SmFeAs(O,F). This
could be ascribed to virtually clean SC gaps\cite{HardyBurger2010}
in the case of Ba(Fe$_{1-x}$Co$_{x}$)$_{2}$As$_{2}$ (and other
122 systems\cite{MerteljKusar2010}) in comparison to SmFeAs(O,F)
where the relaxation dynamics and a large low-$T$ heat capacity\cite{TropeanoMartinelli2008}
suggest the presence of ungapped parts of the Fermi surface in the
SC state.\cite{MerteljKusar2010}

\begin{figure*}
\includegraphics[angle=-90,scale=0.6]{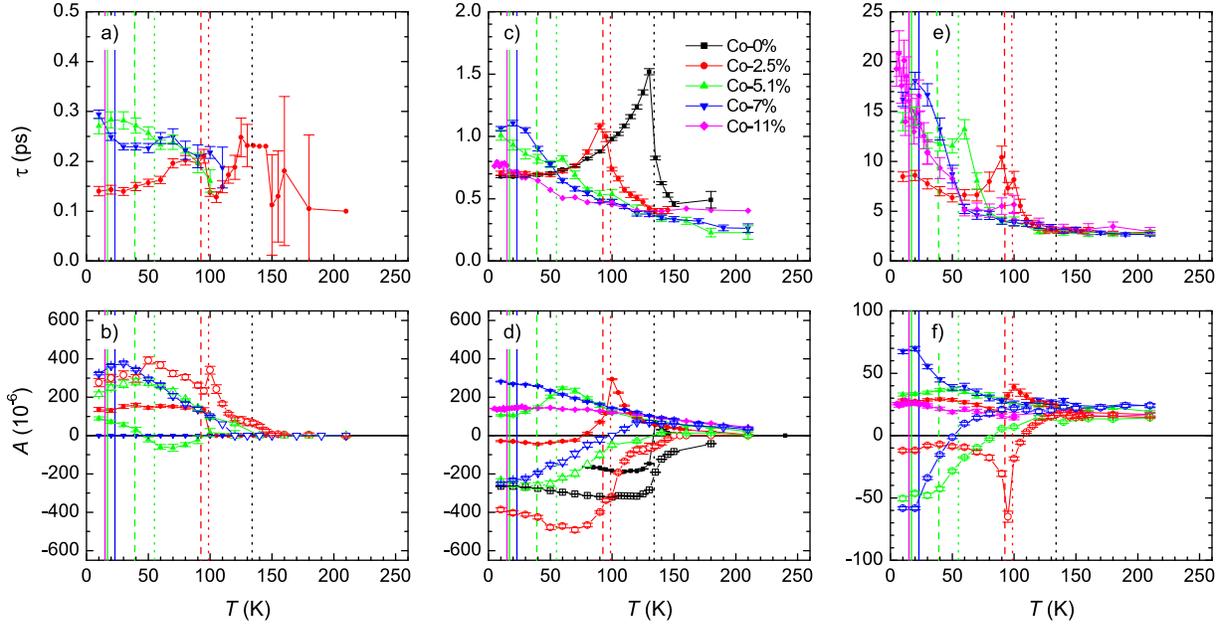}

\caption{(Color online) Exponential relaxation fit relaxation times (a), (c),
(e) and corresponding amplitudes (b), (d), (f). The empty and filled
symbols represent $\mathcal{P}^{+}$ and $\mathcal{P}^{-}$ probe
polarization, respectively. The vertical lines indicate $T\mathrm{_{c}}$
(full lines), $T\mathrm{_{SDW}}$ (dashed lines) and $T\mathrm{_{s}}$
(dotted lines).}
\label{fig:fit-all}
\end{figure*}

\subsection{Orthorhombic SDW state}

\subsubsection{Determination of the relaxation components}

While in the Co-0\% sample at low $T$ the initial relaxation can
be fit with a single exponential decay in both polarizations the samples
with finite Co dopings clearly show a multi-component relaxation.
In order to consistently fit the transients for both probe polarizations
up to 100 ps delay, to determine $T$-dependence of relaxation times,
three exponentially decaying components need to be employed. 

Component A {[}see Fig. \ref{fig:fit-all} (a), (b){]} with relaxation
time $\tau_{\mathrm{A}}\sim0.2$ ps is needed to fit the difference
in the rise time dynamics between the two orthogonal probe polarizations.
The component is absent in the Co-0\% and superconducting Co-11\%
samples and appears below $\sim150$ K in the other samples with the
relaxation time only weakly dependent on the temperature. It could
be associated with the initial relaxation of the high energy optically
excited electrons towards the Fermi energy and/or the inter-band momentum
scattering between states near the Fermi energy at different parts
of the Fermi surface\cite{RettigCortes2010}.

Component B is the main component present in all samples {[}see Fig.
\ref{fig:fit-all} (d){]} representing the initial $0.3-1.5$ ps decay.
The relaxation time, $\tau_{\mathrm{B}}$, {[}see Fig. \ref{fig:fit-all}
(c){]} shows a slowing down%
\footnote{Due to the presence of the surface-strain bias the transition is,
strictly speaking, a crossover.\cite{ChuAnalytis2010}%
} at $T_{\mathrm{s}}$ in the samples with the SDW order. The peak
in $\tau_{\mathrm{B}}$ is only weakly pronounced in the Co-5.1\%
sample, which shows the SDW-SC coexistence, and is completely absent
in the Co-7\% and Co-11\% samples, where we observe a monotonous increase
of $\tau_{\mathrm{B}}$ with decreasing temperature.%
\footnote{The apparent peak of $\tau_{_{\mathrm{B}}}$ at $T_{\mathrm{c}}$
in the Co-7\% sample is due to the appearance of the SC relaxation
component.%
} 

Component C {[}see Fig. \ref{fig:fit-all} (e), (f){]} has the longest
relaxation time ($\tau_{\mathrm{C}}$) spanning from a few ps at 200K
up to 20 ps at 5K in the Co-7\% and Co-11\% samples.%
\footnote{At high temperature it is strongly influenced by the acoustic shock
wave feature around the 10 ps delay.%
} In the Co-2.5\% and Co-5.1\% samples $\tau_{\mathrm{C}}$ also slows
down near $T_{\mathrm{s}}$. The qualitative behavior of this component
is very similar to component B so it is very likely that both components
together are due to a single process with a non-exponential decay
dynamics.

\subsubsection{Analysis of the temperature dependence}

Below $T\mathrm{_{s}}$ the $\Delta R/R$ amplitude shows different
$T$-dependence for the two orthogonal probe polarizations which is
the most clearly pronounced in the Co-2.5\% sample {[}see Fig. \ref{fig:fig-AvsT-2-All}
(a){]}. This indicates that the states involved in the two orthogonal
probe polarizations correspond to different parts of the Fermi surface,%
\footnote{The SC component shows the same $T$-dependence for both polarizations
consistent with isotropic SC gaps.%
} presumably originating from different bands crossing the Fermi energy
as confirmed by the recent ARPES photon-polarization analysis in untwinned
Ba(Fe$_{1-x}$Co$_{x}$)$_{2}$As$_{2}$.\cite{YiLu2011} 

To analyze the anisotropic $T$-dependence we first rewrite equation
(4) from Ref. {[}\onlinecite{DvorsekKabanov2002}{]}, that describes
the photoinduced reflectivity change due to the presence of photoexcited
carriers, in a more general form for a pair of bands:\begin{align}
\Delta R_{\alpha,\beta} & \propto\int\mathrm{d}^{3}k[\left|M_{\alpha,\beta}(\mathbf{k})\right|^{2}\Delta f_{\mathrm{\alpha}}(\mathbf{k})\times\nonumber \\
 & \times g\left(\epsilon_{\beta}(\mathbf{k})-\epsilon_{\alpha}(\mathbf{k})-\hbar\omega_{\mathrm{probe}}\right)].\label{eq:DR}\end{align}
Here $M_{\alpha,\beta}$ is the effective probe-polarization dependent
optical-dipole matrix element between an initial band, $\alpha$,
and a final band, $\beta$, $\Delta f_{\alpha}(\mathbf{k})$ the photoexcited
change of the charge-carrier distribution function in the initial
band, $g(\epsilon)$ the effective transition line-shape and $\hbar\omega_{\mathrm{probe}}$
the probe photon energy. For simplicity we assumed that the energy
of the final band is far from the Fermi energy, $\left|\epsilon_{\beta}(\mathbf{k})-\epsilon_{\mathrm{F}}\right|\sim\hbar\omega_{\mathrm{probe}}\gg k_{\mathrm{B}}T$,
so $\Delta f_{\beta}(\mathbf{k})$ can be neglected after the fast
initial relaxation of the ultra-hot carriers.

Integral (\ref{eq:DR}) selectively samples $\Delta f_{\alpha}(\mathbf{k})$
in different regions of $k$-space depending on the probe polarization
and photon energy. Due to contributions of several optical transitions
with finite effective line-widths it is usually assumed that (\ref{eq:DR})
smoothly samples over the relevant energy range in the vicinity of
the Fermi energy and $\Delta R$ can be approximated by the total
photexcited carrier density, $\Delta R=\gamma n_{\mathrm{pe}}$,\cite{KabanovDemsar99,DvorsekKabanov2002}
and any change of $\Delta R$ upon change of external parameters ($T$
for example) is attributed to the change of $n_{\mathrm{pe}}$ while
the proportionality factor $\gamma$ is assumed to be a constant.

In Ba(Fe$_{1-x}$Co$_{x}$)$_{2}$As$_{2}$ however, a complex band
structure reorganization with bands shifting by as much as 80 meV
has been observed below $T_{\mathrm{s}}$.\cite{YiLu2011} These shifts
can significantly modify the sampling region of the integral (\ref{eq:DR})
and violate the assumption of a constant $\gamma$. To take this into
account we therefore assume that $\gamma$ is temperature dependent
and expand it in terms of an order parameter. The order parameter
can be associated with the opening of a partial $T$-dependent charge
gap $\Delta(T)$ upon the Fermi surface reconstruction below $T\mathrm{_{s}}$.\cite{AnalytisMcDonald2009}
Assuming a complex BCS-like order parameter with the magnitude $\Delta(T)$
below $T_{\mathrm{s}}$ we obtain:\begin{equation}
\mbox{\ensuremath{\Delta}}R=(\gamma_{0}+\eta\Delta(T){}^{2})n_{\mathrm{pe}}.\label{eq:DR-npe}\end{equation}
Since for the $\mathcal{P}^{-}$ probe polarization the $T$-dependent
$\Delta R/R$ amplitude shows the characteristic shape which is associated
with an appearance of a bottleneck in the photo-excited electron relaxation
below $T_{\mathrm{s}}$ we use the bottleneck model from Kabanov \textit{et
al.}\cite{KabanovDemsar99},

\begin{alignat}{1}
n_{\mathrm{pe}}\propto\nicefrac{1}{\left[\left(1+\frac{k_{\mathrm{B}}T}{2\Delta\left(T\right)}\right)\left(1+g_{\mathrm{ph}}\sqrt{\frac{k_{\mathrm{B}}T}{\Delta\left(T\right)}}\exp\left(-\frac{k_{\mathrm{B}}T}{\Delta\left(T\right)}\right)\right)\right],}\label{eq:AvsT}\end{alignat}
to describe $T$-dependence of $n_{\mathrm{pe}}$. Using the BCS temperature
dependent gap we can obtain a good fit of equation (\ref{eq:DR-npe})
to the $\Delta R/R$ amplitude for both probe polarizations (see Fig.
\ref{fig:fig-AvsT-2-All}). The relative gap magnitudes are consistent
(see Table \ref{tbl:gaps}) with previously reported values\cite{StojchevskaKusar2010}
in different iron pnictides and show a decrease with doping, consistent
with a decrease of the stability of the orthorhombic SDW state.

\subsection{Normal state}

\subsubsection{Normal state bottleneck and pseudogap}

Above $T_{\mathrm{s}}$ the $T$-dependent $\Delta R/R$ amplitude
shows tails which can not be described by (\ref{eq:AvsT}). These
tails indicate a bottleneck in relaxation and therefore the presence
of a pseudogap persisting up to $\sim200\mbox{ K}$. With increasing
Co doping the bottleneck becomes even more pronounced at higher $T$
and remains present also in the non-SDW Co-7\% and Co-11\% samples. 

The $\mathcal{F}$ dependence of the $\Delta R/R$ transients shown
in Fig. \ref{fig:fig-DR-vs-F-5.1=000025} (e) and (f) indicates that
at high excitation the relaxation component, which is responsible
for the tails, saturates and the shapes of transients become almost
identical to the room temperature ones. The observed external saturation
fluences of the order of 100 $\mu$J/cm$^{2}$ correspond to the absorbed
optical energy of $\sim50$ K/Fe. This amount of energy would thermally
heat the experimental volume for only a few K. Any property or a state,
that is responsible for the tails, can therefore be non-thermally
destroyed. This rules out any static effect, such as the surface-strain
bias, a rigid band shift with $T$ or a band-structure pseudogap,
for example, as a possible origins of the tails.

To obtain a quantitative information about the pseudogap we analyze
the normal state $T$-dependent $\Delta R/R$ magnitude in the context
of the relaxation across a $T$-independent gap.\cite{KabanovDemsar99,MerteljKabanov2009prl,MerteljKusar2010}
Assuming that in the normal state any $T$-dependence of $\gamma$
can be neglected, we fit the $\mathcal{P}^{+}$ $\Delta R/R$ magnitude%
\footnote{The $T$-dependence of the $\mathcal{P}^{-}$ $\Delta R/R$ magnitude
is qualitatively similar with an offset due to another relaxation
process.%
} by: \begin{alignat}{1}
\mbox{\ensuremath{\Delta}}R\propto & n_{\mathrm{pe}}\propto\left[1+g_{\mathrm{ph}}\exp\left(-\frac{k_{\mathrm{B}}T}{\Delta_{\mathrm{PG}}\left(T\right)}\right)\right]^{-1},\label{eq:AvsT-PG}\end{alignat}
where $g_{\mathrm{ph}}$ is the ratio between the number of involved
phonons and number of involved quasi-particle states.\cite{KabanovDemsar99} 

The obtained pseudogap magnitudes $2\Delta_{\mathrm{PG}}$ (see Table
\ref{tbl:gaps}) are very similar to spin pseudogap magnitudes obtained
form $T$-dependence of the Knight shift\cite{NingAhilan2009} suggesting
that a suppression of density of states in the fluctuation region
is present in both, spin and charge, densities of states. The presence
of the charge pseudogap is supported also by the $T$-dependence of
the $c$-axis electrical resistivity\cite{TanatarThaler2010} and
the V shape of the tunneling conductance spectra\cite{ChuangAllan2010}. 

It should be noted that the charge pseudogap was observed also in
the electron doped SmFeAs(O,F), \cite{MerteljKabanov2009prl,MerteljKusar2010}
which, similarly as Ba(Fe$_{1-x}$Co$_{x}$)$_{2}$As$_{2}$, shows
the spin pseudogap.\cite{AhilanNing2009} 

\begin{table}
\begin{tabular}{c|cc}
$x$ & $\nicefrac{2\Delta(0)}{k_{\mathrm{B}}T\mathrm{_{s}}}$ & $2\Delta_{\mathrm{PG}}$ (K)\tabularnewline
\hline
0\% & $9\pm2$ & -\tabularnewline
2.5\% & $7\pm2$ & -\tabularnewline
5.1\% & $4\pm2$ & $800\pm100$\tabularnewline
7\% & - & $660\pm100$\tabularnewline
11\% & - & $610\pm100$\tabularnewline
\end{tabular}

\caption{Charge gap magnitudes in the SDW samples and characteristic pseudogap
energies in the SC samples as obtained from the fits in Fig. \ref{fig:fig-AvsT-2-All}
discussed in text.}
\label{tbl:gaps}
\end{table}

\subsubsection{Anisotropy and nematic fluctuations}

One of the most striking features of our data set is the observation
of the anisotropy of the optical transients above $T_{\mathrm{s}}$
in the tetragonal phase without any deliberately applied external
uniaxial stress. The absence of the anisotropy at the room temperature
and in the Co-11\% sample proves that the observed anisotropy is not
an experimental artifact but is intrinsic to our samples. At low dopings
the high-$T$ anisotropy axes match the orthorhombic-state anisotropy
axes indicating that also the high-$T$ anisotropy is oriented along
the orthorhombic crystal axes direction. 

In the absence of any structural data which would indicate that our
samples are not tetragonal in the thermodynamic equilibrium above
$T_{\mathrm{s}}$, we assume that the breaking of the 4-fold tetragonal
symmetry is not spontaneous, but is a consequence of anisotropic boundary
and/or excitation conditions that introduce an anisotropic surface
strain. 

We believe that the strain is a consequence of the local crystal expansion
due to the local thermal load. The anisotropy of the strain could
be linked to the unidirectional terraces observed on the surface of
the cleaved crystals. The strain is expected to be weak since the
average increase of the temperature in the experimental volume is
at most a few K. The anisotropic response of the sample is therefore
possible only if the system is very close to a spontaneous symmetry-breaking
instability of the 4-fold point symmetry.

Due to the concurrent appearance of the bottleneck and the polarization
anisotropy of the transients around $\mbox{200 K}$ the pseudogap
might be associated with nematic orbital fluctuations/ordering of
the Fe $d$ orbitals which breaks the 4-fold point symmetry. The ordering
is not necessary static since the timescale of relaxation is below
$\sim0.5$ ps. Similar nematic ordering, albeit static and at somewhat
lower temperatures, was observed also by other techniques.\cite{ChuangAllan2010,ChuAnalytis2010,DuszaLucarelli2011} 

The nematic fluctuations appear to be particularly strong in the Co-5.1\%
and Co-7\% samples where they order due to the surface-strain bias
resulting in the probe polarization anisotropy and the change of the
sign of the $\mathcal{P}^{-}$-probe-polarization transients well
above any transition at $\sim110$K and $\sim70K$ in the Co-5.1\%
and 7\% samples, respectively {[}see Fig. \ref{fig:fig-AvsT-2-All}
(b){]}. 

In the Co-11\% sample no probe polarization anisotropy and therefore
no macroscopic ordering of the nematic fluctuations is observed. However,
the temperature dependence of the $\Delta R/R$ magnitude in this
sample is very similar to the $\Delta R/R$ magnitude of the polarization-averaged
$\Delta R/R$ transients in the Co-7\% sample {[}see Fig. \ref{fig:fig-AvsT-2-All}
(b){]} indicating that the pseudogap region extends well into the
overdoped region of the phase diagram.

\subsection{Electron phonon coupling }

\begin{figure}[tbh]
\begin{centering}
\includegraphics[width=0.45\textwidth]{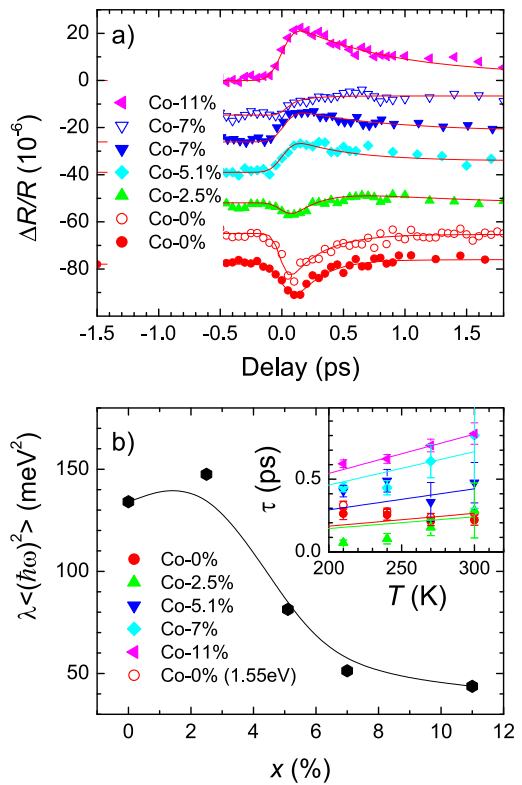} 
\par\end{centering}

\caption{(Color online) The initial part of $\Delta R/R$ transients at 300K
together with exponential fits as a function of Co doping (a). Open
symbols correspond to the transients measured with 1.55 eV pump-photon
energy in the Co-0\% and Co-7\% samples. The second moment of the
Eliashberg function as a function of Co doping (b). The thin line
is an eye-guide. The $T$-dependence of the initial relaxation time
together with fits of Equation (\ref{eq:TauPoorHigh}) is shown in
the inset.}

\label{fig:fig-lw2} 
\end{figure}

\subsubsection{Determination of $\lambda\langle\omega^{2}\rangle$}

At room temperature Ba(Fe$_{1-x}$Co$_{x}$)$_{2}$As$_{2}$ is a
bad metal with resistivity in the sub-m$\Omega$cm range\cite{ChuAnalytis2009}
and plasma frequency in an $\sim1$ eV range\cite{HuDong2008,LucarelliDusza2010}.
Since our data suggest that the effects of the nematic fluctuations
become negligible around room temperature we analyze the initial part
of the $\Delta R/R$ transients at high $T$ in the framework of the
electron-phonon relaxation in metals.\cite{KabanovAlexandrov2008,GadermaierAlexandrov2010}
The $\mathcal{F}$-independent relaxation at the room temperature
{[}see Fig. \ref{fig:fig-DR-vs-F-5.1=000025} (g) and (h){]} warrants
use of the low excitation expansion\cite{KabanovAlexandrov2008},
where in the high temperature limit the energy relaxation time is
proportional to $T$,\cite{KabanovAlexandrov2008,GadermaierAlexandrov2010}\begin{equation}
\tau=\frac{2\pi k_{\mathrm{B}}T}{3\hbar\lambda\langle\omega^{2}\rangle}.\label{eq:TauPoorHigh}\end{equation}
Here $\lambda\langle\omega^{2}\rangle$ is the second moment of the
Eliashberg function, $\alpha^{2}F(\omega)$, and $k_{\mathrm{B}}$
the Boltzman constant.\cite{KabanovAlexandrov2008} The equation is
expected to be valid for $k_{\mathrm{B}}T>$$\hbar\omega_{0}$, where
$\omega_{0}$ is the characteristic Eliashberg-function phonon frequency.
Estimating $\hbar\omega_{0}\sim20$ meV being a half of the maximum
phonon frequency in BaFe$_{2}$As$_{2}$,\cite{MittalSu2008} we expect
(\ref{eq:TauPoorHigh}) to be valid above $\sim200$ K.

We therefore determine the relaxation time above 200 K by fitting
the initial part of the $\Delta R/R$ transients with a finite-rise-time
single exponential decay\cite{MihailovicDemsar99} {[}see Fig. \ref{fig:fig-lw2}
(a){]}.%
\footnote{The Co-2.5\% sample shows a clear two component initial decay at the
room temperature so a two exponential fit was used.%
} The temperature dependence of the relaxation time shows a clear linear
$T$-dependence predicted by equation (\ref{eq:TauPoorHigh}) only
in the near optimally doped Co-7\% and overdoped Co-11\% samples {[}see
inset of Fig. \ref{fig:fig-lw2} (b){]}. At lower Co dopings there
is a clear departure from the linear $T$-dependence below $\sim250$
K indicating that $\omega_{0}$ rises with decreasing Co doping and/or
the effects of the nematic fluctuations can not be neglected up to
$\sim250$ K.

In the non SC samples $\lambda\langle\omega^{2}\rangle$ is similar
as in SrFe$_{2}$As$_{2}$ and SmAsFeO indicating a moderate electron
phonon coupling constant $\lambda\sim0.3$.\cite{StojchevskaKusar2010}
The value of $\lambda\langle\omega^{2}\rangle$ strongly drops in
the superconducting samples consistent with high excitation density
result.\cite{MansartBoschetto2010} The decrease of $\lambda\langle\omega^{2}\rangle$
with increasing Co doping {[}see Fig. \ref{fig:fig-lw2} (b){]} suggests
even lower $\lambda\sim0.1$ in the SC samples, however, due to the
decrease of $\omega_{o}$ the relative decrease of $\lambda$ might
be less than that of $\lambda\langle\omega^{2}\rangle$.

\subsubsection{Possible multi-band effects}

Owing to the multi-band nature of iron-pnictides it is possible, that
(due to optical selection rules) relaxation in some bands with possible
higher couplings is not directly detected in $\Delta R/R$ transients.
This would happen if the inter-band momentum scattering was slower
than the energy relaxation rate in the unobserved band(s). The slow-inter-band-scattering-rate
possibility is suggested by the marked pump-photon energy dispersion
of the Co-7\% sample transients at high $T$ {[}see inset to Fig.
\ref{fig:fig-lw2} (a){]}, which should be absent in the case of a
very fast inter-band-momentum scattering rate.

Trying to clarify this we estimate the upper bound for the inter-band
momentum scattering rate, $\nicefrac{1}{\tau_{\mathrm{IB}}}$, by
the scattering rate of the narrow Drude peak.\cite{LucarelliDusza2010}
From Ref. {[}\onlinecite{LucarelliDusza2010}{]} we obtain $\tau_{\mathrm{IB}}\gtrsim\nicefrac{1}{30\mathrm{\, cm}^{-1}}\sim200$
fs which is indeed comparable to the measured energy relaxation time
{[}Fig. \ref{fig:fig-lw2} (b){]} indicating a possibility that the
estimated $\lambda\langle\omega^{2}\rangle$ is not well averaged
over different bands. On the other hand, different analyses of the
optical conductivity,\cite{HuDong2008,TuLi2010,NakajimaIshida2010,BarisicWu2010}
albeit in samples of different origin than ours,%
\footnote{Our samples have the same origin as in Ref. {[}\onlinecite{LucarelliDusza2010}{]}
and $\nicefrac{1}{\tau_{_{\mathrm{IB}}}}$ can be strongly affected
by impurities.%
} result in a much broader Drude peak and consequently in at least
ten times shorter $\tau_{\mathrm{IB}}$ therefore supporting the fast
inter-band-momentum scattering rate scenario. 

The observed pump-photon energy dispersion of the transients {[}see
inset to Fig. \ref{fig:fig-lw2} (a){]}, which is negligible in the
Co-0\% sample, therefore suggests, that $\lambda\langle\omega{}^{2}\rangle$
is well averaged over the bands in the undoped sample. This can not
be claimed (due to the significant pump-photon energy dispersion in
the Co-7\% sample) for the samples with the Co doping well in the
SC dome. Preliminary experiments with different pump/probe photon
energies and better time resolution have, however, so far not shown
the presence of any additional faster relaxation component and consequently
a presence of bands with larger $\lambda\langle\omega{}^{2}\rangle$
for any of the present Co doping levels.%
\footnote{C. Gadermaier \emph{et al.}, unpublished data.%
}

\section{Summary and conclusions}

We investigated doping dependence of electronic properties in Ba(Fe$_{1-x}$Co$_{x}$)$_{2}$As$_{2}$
by means of time resolved optical pump-probe spectroscopy. We observe
a smooth evolution of the response with the Co doping as the system
crosses over from the undoped SDW ground state to the superconducting
ground state. 

In the undoped and underdoped samples ($x\lesssim5.1\%$) a clear
signature of a bottleneck formation in the relaxation of the photoexcited
QP is observed below the tetragonal to orthorhombic structural transition
temperature $T\mathrm{_{s}}$. The bottleneck is attributed to the
partial charge gap opening due to the band-structure reconstruction
below $T\mathrm{_{s}}$, similar to other undoped iron pnictides.\cite{StojchevskaKusar2010}
The relative charge gap magnitude $\nicefrac{2\Delta(0)}{k_{\mathrm{B}}T_{\mathrm{s}}}$
decreases with Co doping, consistent with a decrease of the stability
of the orthorhombic/SDW state. 

Similar to SmFeAs(O,F) we observe an anomalous $T$ dependence of
the relaxation in the normal state with the addition of the 2-fold
anisotropy in the tetragonal state. Although we were not able to determine
the precise origin of the observed 2-fold anisotropy, the observation
of unidirectional terraces on the surface of the cleaved crystals
suggests, that this anisotropy might be due to the anisotropic component
of the surface strain induced by the laser heating. Since this strain
is weak this indicates a high nematic susceptibility of the samples.
The anomalous normal state behavior is therefore attributed to electronic
nematic fluctuations, that persist up to $\sim200$ K, and open a
pseudogap in the density of states near the Fermi energy. The fluctuation
region extends well above $T_{\mathrm{s}}$ in the underdoped region
and all over the SC dome region of the phase diagram. Due to the surface-strain
bias these fluctuations tend to align resulting in an anisotropic
optical response also in the tetragonal near-optimally-doped 7\%-Co
sample.

Surprisingly, no clear separate anomaly is observed upon SDW formation
at slightly lower $T_{\mathrm{SDW}}$ in underdoped samples ($x=2.5\%$
and 5.1\%). This suggests that the mechanism responsible for the pseudogap
formation, nematic orbital fluctuations and partial gap opening below
$T_{\mathrm{s}}$ is, despite the ubiquitous coupling to spins, not
spin driven.

At room temperature, where the nematic fluctuations are negligible,
the transients are analyzed in the framework of the Fermi-liquid electron-phonon
relaxation model.\cite{KabanovAlexandrov2008} The analysis indicates,
as in SrFe$_{2}$As$_{2}$ and SmAsFeO,\cite{StojchevskaKusar2010}
a moderate electron phonon coupling. The second moment of the Eliashberg
function $\lambda<\omega^{2}>$ is found to decrease with the Co doping
resulting in a decrease of estimated $\lambda$ from$\sim0.3$ in
the nonsuperconducting to $\sim0.1$ in the superconducting samples.
It is not clear however, to what extent at higher Co dopings the systematic
error due to a possible slow inter-band momentum scattering contributes
to this decrease.

In the SC state an additional relaxation component appears in the
$\Delta R/R$ transients. The behavior of the SC component is consistent
with isotropic SC gaps, that have the BCS $T$ dependence. The amplitude
of the SC component saturates with increasing excitation density.
The saturation is associated with a complete non-thermal destruction
of the SC state, which proceeds on a sub 0.5-ps timescale.

In the near optimally doped sample with 7\% Co doping ($T_{\mathrm{c}}\sim23$
K) the determined SC state optical destruction energy density, $U_{\mathrm{p}}/k_{\mathrm{B}}=0.3$
K/Fe, is twice the thermodynamic condensation energy. A half of the
deposited optical energy is therefore transferred to the low frequency
non-pair-breaking phonons on a sub-0.5-ps timescale. Comparison with
SmFeAs(O,F)\cite{MerteljKusar2010} ($T_{\mathrm{c}}\sim49$ K) indicates
that $U_{\mathrm{p}}$ roughly scales as $T_{\mathrm{c}}^{2}$. The
SC state recovery dynamics in Ba(Fe$_{1-x}$Co$_{x}$)$_{2}$As$_{2}$
is slower than in SmFeAs(O,F) suggesting, contrary to SmFeAs(O,F),
clean SC gaps in Ba(Fe$_{1-x}$Co$_{x}$)$_{2}$As$_{2}$.
\begin{acknowledgments}
Work at Jozef Stefan Institute was supported by ARRS (Grant No. P1-0040).
Work at Stanford University was supported by the Department of Energy,
Office of Basic Energy Sciences under contract DE-AC02-76SF00515. 

We would like to thank M. Strojnik for help with AFM surface characterization
and V.V. Kabanov for fruitful discussions.
\end{acknowledgments}
\bibliography{biblio}

\begin{thebibliography}{65}%
\makeatletter
\providecommand \@ifxundefined [1]{%
 \@ifx{#1\undefined}
}%
\providecommand \@ifnum [1]{%
 \ifnum #1\expandafter \@firstoftwo
 \else \expandafter \@secondoftwo
 \fi
}%
\providecommand \@ifx [1]{%
 \ifx #1\expandafter \@firstoftwo
 \else \expandafter \@secondoftwo
 \fi
}%
\providecommand \natexlab [1]{#1}%
\providecommand \enquote  [1]{``#1''}%
\providecommand \bibnamefont  [1]{#1}%
\providecommand \bibfnamefont [1]{#1}%
\providecommand \citenamefont [1]{#1}%
\providecommand \href@noop [0]{\@secondoftwo}%
\providecommand \href [0]{\begingroup \@sanitize@url \@href}%
\providecommand \@href[1]{\@@startlink{#1}\@@href}%
\providecommand \@@href[1]{\endgroup#1\@@endlink}%
\providecommand \@sanitize@url [0]{\catcode `\\12\catcode `\$12\catcode
  `\&12\catcode `\#12\catcode `\^12\catcode `\_12\catcode `\%12\relax}%
\providecommand \@@startlink[1]{}%
\providecommand \@@endlink[0]{}%
\providecommand \url  [0]{\begingroup\@sanitize@url \@url }%
\providecommand \@url [1]{\endgroup\@href {#1}{\urlprefix }}%
\providecommand \urlprefix  [0]{URL }%
\providecommand \Eprint [0]{\href }%
\providecommand \doibase [0]{http://dx.doi.org/}%
\providecommand \selectlanguage [0]{\@gobble}%
\providecommand \bibinfo  [0]{\@secondoftwo}%
\providecommand \bibfield  [0]{\@secondoftwo}%
\providecommand \translation [1]{[#1]}%
\providecommand \BibitemOpen [0]{}%
\providecommand \bibitemStop [0]{}%
\providecommand \bibitemNoStop [0]{.\EOS\space}%
\providecommand \EOS [0]{\spacefactor3000\relax}%
\providecommand \BibitemShut  [1]{\csname bibitem#1\endcsname}%
\let\auto@bib@innerbib\@empty
%</preamble>
\bibitem [{\citenamefont {Kamihara}\ \emph {et~al.}(2006)\citenamefont
  {Kamihara}, \citenamefont {Hiramatsu}, \citenamefont {Hirano}, \citenamefont
  {Kawamura}, \citenamefont {Yanagi}, \citenamefont {Kamiya},\ and\
  \citenamefont {Hosono}}]{KamiharaKamihara2006}%
  \BibitemOpen
  \bibfield  {author} {\bibinfo {author} {\bibfnamefont {Y.}~\bibnamefont
  {Kamihara}}, \bibinfo {author} {\bibfnamefont {H.}~\bibnamefont {Hiramatsu}},
  \bibinfo {author} {\bibfnamefont {M.}~\bibnamefont {Hirano}}, \bibinfo
  {author} {\bibfnamefont {R.}~\bibnamefont {Kawamura}}, \bibinfo {author}
  {\bibfnamefont {H.}~\bibnamefont {Yanagi}}, \bibinfo {author} {\bibfnamefont
  {T.}~\bibnamefont {Kamiya}}, \ and\ \bibinfo {author} {\bibfnamefont
  {H.}~\bibnamefont {Hosono}},\ }\href {\doibase 10.1021/ja063355c} {\bibfield
  {journal} {\bibinfo  {journal} {Journal of the American Chemical Society}\
  }\textbf {\bibinfo {volume} {128}},\ \bibinfo {pages} {10012} (\bibinfo
  {year} {2006})}\BibitemShut {NoStop}%
\bibitem [{\citenamefont {Kamihara}\ \emph {et~al.}(2008)\citenamefont
  {Kamihara}, \citenamefont {Watanabe}, \citenamefont {Hirano}, \citenamefont
  {Hosono} \emph {et~al.}}]{kamiharaWatanabe2008}%
  \BibitemOpen
  \bibfield  {author} {\bibinfo {author} {\bibfnamefont {Y.}~\bibnamefont
  {Kamihara}}, \bibinfo {author} {\bibfnamefont {T.}~\bibnamefont {Watanabe}},
  \bibinfo {author} {\bibfnamefont {M.}~\bibnamefont {Hirano}}, \bibinfo
  {author} {\bibfnamefont {H.}~\bibnamefont {Hosono}},  \emph {et~al.},\
  }\href@noop {} {\bibfield  {journal} {\bibinfo  {journal} {J. Am. Chem. Soc}\
  }\textbf {\bibinfo {volume} {130}},\ \bibinfo {pages} {3296} (\bibinfo {year}
  {2008})}\BibitemShut {NoStop}%
\bibitem [{\citenamefont {Ren}\ \emph {et~al.}(2008)\citenamefont {Ren},
  \citenamefont {Che}, \citenamefont {Dong}, \citenamefont {Yang},
  \citenamefont {Lu}, \citenamefont {Yi}, \citenamefont {Shen}, \citenamefont
  {Li}, \citenamefont {Sun}, \citenamefont {Zhou} \emph {et~al.}}]{RenChe2008}%
  \BibitemOpen
  \bibfield  {author} {\bibinfo {author} {\bibfnamefont {Z.}~\bibnamefont
  {Ren}}, \bibinfo {author} {\bibfnamefont {G.}~\bibnamefont {Che}}, \bibinfo
  {author} {\bibfnamefont {X.}~\bibnamefont {Dong}}, \bibinfo {author}
  {\bibfnamefont {J.}~\bibnamefont {Yang}}, \bibinfo {author} {\bibfnamefont
  {W.}~\bibnamefont {Lu}}, \bibinfo {author} {\bibfnamefont {W.}~\bibnamefont
  {Yi}}, \bibinfo {author} {\bibfnamefont {X.}~\bibnamefont {Shen}}, \bibinfo
  {author} {\bibfnamefont {Z.}~\bibnamefont {Li}}, \bibinfo {author}
  {\bibfnamefont {L.}~\bibnamefont {Sun}}, \bibinfo {author} {\bibfnamefont
  {F.}~\bibnamefont {Zhou}},  \emph {et~al.},\ }\href@noop {} {\bibfield
  {journal} {\bibinfo  {journal} {EPL-Europhysics Letters}\ }\textbf {\bibinfo
  {volume} {83}},\ \bibinfo {pages} {17002} (\bibinfo {year}
  {2008})}\BibitemShut {NoStop}%
\bibitem [{\citenamefont {Liu}\ \emph {et~al.}(2008{\natexlab{a}})\citenamefont
  {Liu}, \citenamefont {Wu}, \citenamefont {Wu}, \citenamefont {Fang},
  \citenamefont {Chen}, \citenamefont {Li}, \citenamefont {Liu}, \citenamefont
  {Xie}, \citenamefont {Wang}, \citenamefont {Yang}, \citenamefont {Ding},
  \citenamefont {He}, \citenamefont {Feng},\ and\ \citenamefont
  {Chen}}]{LiuWu2008}%
  \BibitemOpen
  \bibfield  {author} {\bibinfo {author} {\bibfnamefont {R.~H.}\ \bibnamefont
  {Liu}}, \bibinfo {author} {\bibfnamefont {G.}~\bibnamefont {Wu}}, \bibinfo
  {author} {\bibfnamefont {T.}~\bibnamefont {Wu}}, \bibinfo {author}
  {\bibfnamefont {D.~F.}\ \bibnamefont {Fang}}, \bibinfo {author}
  {\bibfnamefont {H.}~\bibnamefont {Chen}}, \bibinfo {author} {\bibfnamefont
  {S.~Y.}\ \bibnamefont {Li}}, \bibinfo {author} {\bibfnamefont
  {K.}~\bibnamefont {Liu}}, \bibinfo {author} {\bibfnamefont {Y.~L.}\
  \bibnamefont {Xie}}, \bibinfo {author} {\bibfnamefont {X.~F.}\ \bibnamefont
  {Wang}}, \bibinfo {author} {\bibfnamefont {R.~L.}\ \bibnamefont {Yang}},
  \bibinfo {author} {\bibfnamefont {L.}~\bibnamefont {Ding}}, \bibinfo {author}
  {\bibfnamefont {C.}~\bibnamefont {He}}, \bibinfo {author} {\bibfnamefont
  {D.~L.}\ \bibnamefont {Feng}}, \ and\ \bibinfo {author} {\bibfnamefont
  {X.~H.}\ \bibnamefont {Chen}},\ }\href {\doibase
  10.1103/PhysRevLett.101.087001} {\bibfield  {journal} {\bibinfo  {journal}
  {Phys. Rev. Lett.}\ }\textbf {\bibinfo {volume} {101}},\ \bibinfo {pages}
  {087001} (\bibinfo {year} {2008}{\natexlab{a}})}\BibitemShut {NoStop}%
\bibitem [{\citenamefont {Ahilan}\ \emph {et~al.}(2008)\citenamefont {Ahilan},
  \citenamefont {Ning}, \citenamefont {Imai}, \citenamefont {Sefat},
  \citenamefont {Jin}, \citenamefont {McGuire}, \citenamefont {Sales},\ and\
  \citenamefont {Mandrus}}]{AhilanNing2009}%
  \BibitemOpen
  \bibfield  {author} {\bibinfo {author} {\bibfnamefont {K.}~\bibnamefont
  {Ahilan}}, \bibinfo {author} {\bibfnamefont {F.~L.}\ \bibnamefont {Ning}},
  \bibinfo {author} {\bibfnamefont {T.}~\bibnamefont {Imai}}, \bibinfo {author}
  {\bibfnamefont {A.~S.}\ \bibnamefont {Sefat}}, \bibinfo {author}
  {\bibfnamefont {R.}~\bibnamefont {Jin}}, \bibinfo {author} {\bibfnamefont
  {M.~A.}\ \bibnamefont {McGuire}}, \bibinfo {author} {\bibfnamefont {B.~C.}\
  \bibnamefont {Sales}}, \ and\ \bibinfo {author} {\bibfnamefont
  {D.}~\bibnamefont {Mandrus}},\ }\href {\doibase 10.1103/PhysRevB.78.100501}
  {\bibfield  {journal} {\bibinfo  {journal} {Physical Review B (Condensed
  Matter and Materials Physics)}\ }\textbf {\bibinfo {volume} {78}},\ \bibinfo
  {eid} {100501} (\bibinfo {year} {2008})}\BibitemShut {NoStop}%
\bibitem [{\citenamefont {Ning}\ \emph {et~al.}(2009)\citenamefont {Ning},
  \citenamefont {Ahilan}, \citenamefont {Imai}, \citenamefont {Sefat},
  \citenamefont {Jin}, \citenamefont {McGuire}, \citenamefont {Sales},\ and\
  \citenamefont {Mandrus}}]{NingAhilan2009}%
  \BibitemOpen
  \bibfield  {author} {\bibinfo {author} {\bibfnamefont {F.}~\bibnamefont
  {Ning}}, \bibinfo {author} {\bibfnamefont {K.}~\bibnamefont {Ahilan}},
  \bibinfo {author} {\bibfnamefont {T.}~\bibnamefont {Imai}}, \bibinfo {author}
  {\bibfnamefont {A.~S.}\ \bibnamefont {Sefat}}, \bibinfo {author}
  {\bibfnamefont {R.}~\bibnamefont {Jin}}, \bibinfo {author} {\bibfnamefont
  {M.~A.}\ \bibnamefont {McGuire}}, \bibinfo {author} {\bibfnamefont {B.~C.}\
  \bibnamefont {Sales}}, \ and\ \bibinfo {author} {\bibfnamefont
  {D.}~\bibnamefont {Mandrus}},\ }\href {\doibase 10.1143/JPSJ.78.013711}
  {\bibfield  {journal} {\bibinfo  {journal} {Journal of the Physical Society
  of Japan}\ }\textbf {\bibinfo {volume} {78}},\ \bibinfo {pages} {013711}
  (\bibinfo {year} {2009})}\BibitemShut {NoStop}%
\bibitem [{\citenamefont {Hess}\ \emph {et~al.}(2009)\citenamefont {Hess},
  \citenamefont {Kondrat}, \citenamefont {Narduzzo}, \citenamefont
  {Hamann-Borrero}, \citenamefont {Klingeler}, \citenamefont {Werner},
  \citenamefont {Behr},\ and\ \citenamefont {Büchner}}]{HessKondrat2009}%
  \BibitemOpen
  \bibfield  {author} {\bibinfo {author} {\bibfnamefont {C.}~\bibnamefont
  {Hess}}, \bibinfo {author} {\bibfnamefont {A.}~\bibnamefont {Kondrat}},
  \bibinfo {author} {\bibfnamefont {A.}~\bibnamefont {Narduzzo}}, \bibinfo
  {author} {\bibfnamefont {J.~E.}\ \bibnamefont {Hamann-Borrero}}, \bibinfo
  {author} {\bibfnamefont {R.}~\bibnamefont {Klingeler}}, \bibinfo {author}
  {\bibfnamefont {J.}~\bibnamefont {Werner}}, \bibinfo {author} {\bibfnamefont
  {G.}~\bibnamefont {Behr}}, \ and\ \bibinfo {author} {\bibfnamefont
  {B.}~\bibnamefont {Büchner}},\ }\href
  {http://stacks.iop.org/0295-5075/87/i=1/a=17005} {\bibfield  {journal}
  {\bibinfo  {journal} {EPL (Europhysics Letters)}\ }\textbf {\bibinfo {volume}
  {87}},\ \bibinfo {pages} {17005} (\bibinfo {year} {2009})}\BibitemShut
  {NoStop}%
\bibitem [{\citenamefont {Mertelj}\ \emph
  {et~al.}(2009{\natexlab{a}})\citenamefont {Mertelj}, \citenamefont {Kabanov},
  \citenamefont {Gadermaier}, \citenamefont {Zhigadlo}, \citenamefont
  {Katrych}, \citenamefont {Karpinski},\ and\ \citenamefont
  {Mihailovic}}]{MerteljKabanov2009prl}%
  \BibitemOpen
  \bibfield  {author} {\bibinfo {author} {\bibfnamefont {T.}~\bibnamefont
  {Mertelj}}, \bibinfo {author} {\bibfnamefont {V.}~\bibnamefont {Kabanov}},
  \bibinfo {author} {\bibfnamefont {C.}~\bibnamefont {Gadermaier}}, \bibinfo
  {author} {\bibfnamefont {N.}~\bibnamefont {Zhigadlo}}, \bibinfo {author}
  {\bibfnamefont {S.}~\bibnamefont {Katrych}}, \bibinfo {author} {\bibfnamefont
  {J.}~\bibnamefont {Karpinski}}, \ and\ \bibinfo {author} {\bibfnamefont
  {D.}~\bibnamefont {Mihailovic}},\ }\href@noop {} {\bibfield  {journal}
  {\bibinfo  {journal} {Physical Review Letters}\ }\textbf {\bibinfo {volume}
  {102}},\ \bibinfo {pages} {117002} (\bibinfo {year}
  {2009}{\natexlab{a}})}\BibitemShut {NoStop}%
\bibitem [{\citenamefont {Chu}\ \emph {et~al.}(2010)\citenamefont {Chu},
  \citenamefont {Analytis}, \citenamefont {De~Greve}, \citenamefont {McMahon},
  \citenamefont {Islam}, \citenamefont {Yamamoto},\ and\ \citenamefont
  {Fisher}}]{ChuAnalytis2010}%
  \BibitemOpen
  \bibfield  {author} {\bibinfo {author} {\bibfnamefont {J.-H.}\ \bibnamefont
  {Chu}}, \bibinfo {author} {\bibfnamefont {J.~G.}\ \bibnamefont {Analytis}},
  \bibinfo {author} {\bibfnamefont {K.}~\bibnamefont {De~Greve}}, \bibinfo
  {author} {\bibfnamefont {P.~L.}\ \bibnamefont {McMahon}}, \bibinfo {author}
  {\bibfnamefont {Z.}~\bibnamefont {Islam}}, \bibinfo {author} {\bibfnamefont
  {Y.}~\bibnamefont {Yamamoto}}, \ and\ \bibinfo {author} {\bibfnamefont
  {I.~R.}\ \bibnamefont {Fisher}},\ }\href {\doibase 10.1126/science.1190482}
  {\bibfield  {journal} {\bibinfo  {journal} {Science}\ }\textbf {\bibinfo
  {volume} {329}},\ \bibinfo {pages} {824} (\bibinfo {year}
  {2010})}\BibitemShut {NoStop}%
\bibitem [{\citenamefont {Chuang}\ \emph {et~al.}(2010)\citenamefont {Chuang},
  \citenamefont {Allan}, \citenamefont {Lee}, \citenamefont {Xie},
  \citenamefont {Ni}, \citenamefont {Bud'ko}, \citenamefont {Boebinger},
  \citenamefont {Canfield},\ and\ \citenamefont {Davis}}]{ChuangAllan2010}%
  \BibitemOpen
  \bibfield  {author} {\bibinfo {author} {\bibfnamefont {T.-M.}\ \bibnamefont
  {Chuang}}, \bibinfo {author} {\bibfnamefont {M.~P.}\ \bibnamefont {Allan}},
  \bibinfo {author} {\bibfnamefont {J.}~\bibnamefont {Lee}}, \bibinfo {author}
  {\bibfnamefont {Y.}~\bibnamefont {Xie}}, \bibinfo {author} {\bibfnamefont
  {N.}~\bibnamefont {Ni}}, \bibinfo {author} {\bibfnamefont {S.~L.}\
  \bibnamefont {Bud'ko}}, \bibinfo {author} {\bibfnamefont {G.~S.}\
  \bibnamefont {Boebinger}}, \bibinfo {author} {\bibfnamefont {P.~C.}\
  \bibnamefont {Canfield}}, \ and\ \bibinfo {author} {\bibfnamefont {J.~C.}\
  \bibnamefont {Davis}},\ }\href
  {http://www.sciencemag.org/content/327/5962/181.abstract} {\bibfield
  {journal} {\bibinfo  {journal} {Science}\ }\textbf {\bibinfo {volume}
  {327}},\ \bibinfo {pages} {181} (\bibinfo {year} {2010})}\BibitemShut
  {NoStop}%
\bibitem [{\citenamefont {Tanatar}\ \emph
  {et~al.}(2010{\natexlab{a}})\citenamefont {Tanatar}, \citenamefont {Ni},
  \citenamefont {Thaler}, \citenamefont {Bud'ko}, \citenamefont {Canfield},\
  and\ \citenamefont {Prozorov}}]{TanatarThaler2010}%
  \BibitemOpen
  \bibfield  {author} {\bibinfo {author} {\bibfnamefont {M.~A.}\ \bibnamefont
  {Tanatar}}, \bibinfo {author} {\bibfnamefont {N.}~\bibnamefont {Ni}},
  \bibinfo {author} {\bibfnamefont {A.}~\bibnamefont {Thaler}}, \bibinfo
  {author} {\bibfnamefont {S.~L.}\ \bibnamefont {Bud'ko}}, \bibinfo {author}
  {\bibfnamefont {P.~C.}\ \bibnamefont {Canfield}}, \ and\ \bibinfo {author}
  {\bibfnamefont {R.}~\bibnamefont {Prozorov}},\ }\href {\doibase
  10.1103/PhysRevB.82.134528} {\bibfield  {journal} {\bibinfo  {journal} {Phys.
  Rev. B}\ }\textbf {\bibinfo {volume} {82}},\ \bibinfo {pages} {134528}
  (\bibinfo {year} {2010}{\natexlab{a}})}\BibitemShut {NoStop}%
\bibitem [{\citenamefont {{Dusza, A.}}\ \emph {et~al.}(2011)\citenamefont
  {{Dusza, A.}}, \citenamefont {{Lucarelli, A.}}, \citenamefont {{Pfuner, F.}},
  \citenamefont {{Chu, J.-H.}}, \citenamefont {{Fisher, I. R.}},\ and\
  \citenamefont {{Degiorgi, L.}}}]{DuszaLucarelli2011}%
  \BibitemOpen
  \bibfield  {author} {\bibinfo {author} {\bibnamefont {{Dusza, A.}}}, \bibinfo
  {author} {\bibnamefont {{Lucarelli, A.}}}, \bibinfo {author} {\bibnamefont
  {{Pfuner, F.}}}, \bibinfo {author} {\bibnamefont {{Chu, J.-H.}}}, \bibinfo
  {author} {\bibnamefont {{Fisher, I. R.}}}, \ and\ \bibinfo {author}
  {\bibnamefont {{Degiorgi, L.}}},\ }\href {\doibase
  10.1209/0295-5075/93/37002} {\bibfield  {journal} {\bibinfo  {journal} {EPL}\
  }\textbf {\bibinfo {volume} {93}},\ \bibinfo {pages} {37002} (\bibinfo {year}
  {2011})}\BibitemShut {NoStop}%
\bibitem [{\citenamefont {Tanatar}\ \emph
  {et~al.}(2010{\natexlab{b}})\citenamefont {Tanatar}, \citenamefont
  {Blomberg}, \citenamefont {Kreyssig}, \citenamefont {Kim}, \citenamefont
  {Ni}, \citenamefont {Thaler}, \citenamefont {Bud'ko}, \citenamefont
  {Canfield}, \citenamefont {Goldman}, \citenamefont {Mazin},\ and\
  \citenamefont {Prozorov}}]{TanatarBlomberg2010}%
  \BibitemOpen
  \bibfield  {author} {\bibinfo {author} {\bibfnamefont {M.~A.}\ \bibnamefont
  {Tanatar}}, \bibinfo {author} {\bibfnamefont {E.~C.}\ \bibnamefont
  {Blomberg}}, \bibinfo {author} {\bibfnamefont {A.}~\bibnamefont {Kreyssig}},
  \bibinfo {author} {\bibfnamefont {M.~G.}\ \bibnamefont {Kim}}, \bibinfo
  {author} {\bibfnamefont {N.}~\bibnamefont {Ni}}, \bibinfo {author}
  {\bibfnamefont {A.}~\bibnamefont {Thaler}}, \bibinfo {author} {\bibfnamefont
  {S.~L.}\ \bibnamefont {Bud'ko}}, \bibinfo {author} {\bibfnamefont {P.~C.}\
  \bibnamefont {Canfield}}, \bibinfo {author} {\bibfnamefont {A.~I.}\
  \bibnamefont {Goldman}}, \bibinfo {author} {\bibfnamefont {I.~I.}\
  \bibnamefont {Mazin}}, \ and\ \bibinfo {author} {\bibfnamefont
  {R.}~\bibnamefont {Prozorov}},\ }\href {\doibase 10.1103/PhysRevB.81.184508}
  {\bibfield  {journal} {\bibinfo  {journal} {Phys. Rev. B}\ }\textbf {\bibinfo
  {volume} {81}},\ \bibinfo {pages} {184508} (\bibinfo {year}
  {2010}{\natexlab{b}})}\BibitemShut {NoStop}%
\bibitem [{\citenamefont {Ying}\ \emph {et~al.}(2011)\citenamefont {Ying},
  \citenamefont {Wang}, \citenamefont {Wu}, \citenamefont {Xiang},
  \citenamefont {Liu}, \citenamefont {Yan}, \citenamefont {Wang}, \citenamefont
  {Zhang}, \citenamefont {Ye}, \citenamefont {Cheng}, \citenamefont {Hu},\ and\
  \citenamefont {Chen}}]{YingWang2011}%
  \BibitemOpen
  \bibfield  {author} {\bibinfo {author} {\bibfnamefont {J.~J.}\ \bibnamefont
  {Ying}}, \bibinfo {author} {\bibfnamefont {X.~F.}\ \bibnamefont {Wang}},
  \bibinfo {author} {\bibfnamefont {T.}~\bibnamefont {Wu}}, \bibinfo {author}
  {\bibfnamefont {Z.~J.}\ \bibnamefont {Xiang}}, \bibinfo {author}
  {\bibfnamefont {R.~H.}\ \bibnamefont {Liu}}, \bibinfo {author} {\bibfnamefont
  {Y.~J.}\ \bibnamefont {Yan}}, \bibinfo {author} {\bibfnamefont {A.~F.}\
  \bibnamefont {Wang}}, \bibinfo {author} {\bibfnamefont {M.}~\bibnamefont
  {Zhang}}, \bibinfo {author} {\bibfnamefont {G.~J.}\ \bibnamefont {Ye}},
  \bibinfo {author} {\bibfnamefont {P.}~\bibnamefont {Cheng}}, \bibinfo
  {author} {\bibfnamefont {J.~P.}\ \bibnamefont {Hu}}, \ and\ \bibinfo {author}
  {\bibfnamefont {X.~H.}\ \bibnamefont {Chen}},\ }\href {\doibase
  10.1103/PhysRevLett.107.067001} {\bibfield  {journal} {\bibinfo  {journal}
  {Phys. Rev. Lett.}\ }\textbf {\bibinfo {volume} {107}},\ \bibinfo {pages}
  {067001} (\bibinfo {year} {2011})}\BibitemShut {NoStop}%
\bibitem [{\citenamefont {Yi}\ \emph {et~al.}(2011)\citenamefont {Yi},
  \citenamefont {Lu}, \citenamefont {Chu}, \citenamefont {Analytis},
  \citenamefont {Sorini}, \citenamefont {Kemper}, \citenamefont {Moritz},
  \citenamefont {Mo}, \citenamefont {Moore}, \citenamefont {Hashimoto},
  \citenamefont {Lee}, \citenamefont {Hussain}, \citenamefont {Devereaux},
  \citenamefont {Fisher},\ and\ \citenamefont {Shen}}]{YiLu2011}%
  \BibitemOpen
  \bibfield  {author} {\bibinfo {author} {\bibfnamefont {M.}~\bibnamefont
  {Yi}}, \bibinfo {author} {\bibfnamefont {D.}~\bibnamefont {Lu}}, \bibinfo
  {author} {\bibfnamefont {J.-H.}\ \bibnamefont {Chu}}, \bibinfo {author}
  {\bibfnamefont {J.~G.}\ \bibnamefont {Analytis}}, \bibinfo {author}
  {\bibfnamefont {A.~P.}\ \bibnamefont {Sorini}}, \bibinfo {author}
  {\bibfnamefont {A.~F.}\ \bibnamefont {Kemper}}, \bibinfo {author}
  {\bibfnamefont {B.}~\bibnamefont {Moritz}}, \bibinfo {author} {\bibfnamefont
  {S.-K.}\ \bibnamefont {Mo}}, \bibinfo {author} {\bibfnamefont {R.~G.}\
  \bibnamefont {Moore}}, \bibinfo {author} {\bibfnamefont {M.}~\bibnamefont
  {Hashimoto}}, \bibinfo {author} {\bibfnamefont {W.-S.}\ \bibnamefont {Lee}},
  \bibinfo {author} {\bibfnamefont {Z.}~\bibnamefont {Hussain}}, \bibinfo
  {author} {\bibfnamefont {T.~P.}\ \bibnamefont {Devereaux}}, \bibinfo {author}
  {\bibfnamefont {I.~R.}\ \bibnamefont {Fisher}}, \ and\ \bibinfo {author}
  {\bibfnamefont {Z.-X.}\ \bibnamefont {Shen}},\ }\href {\doibase
  10.1073/pnas.1015572108} {\bibfield  {journal} {\bibinfo  {journal}
  {Proceedings of the National Academy of Sciences}\ }\textbf {\bibinfo
  {volume} {108}},\ \bibinfo {pages} {6878} (\bibinfo {year}
  {2011})}\BibitemShut {NoStop}%
\bibitem [{\citenamefont {Rullier-Albenque}\ \emph {et~al.}(2009)\citenamefont
  {Rullier-Albenque}, \citenamefont {Colson}, \citenamefont {Forget},\ and\
  \citenamefont {Alloul}}]{Rullier-AlbenqueColson2009}%
  \BibitemOpen
  \bibfield  {author} {\bibinfo {author} {\bibfnamefont {F.}~\bibnamefont
  {Rullier-Albenque}}, \bibinfo {author} {\bibfnamefont {D.}~\bibnamefont
  {Colson}}, \bibinfo {author} {\bibfnamefont {A.}~\bibnamefont {Forget}}, \
  and\ \bibinfo {author} {\bibfnamefont {H.}~\bibnamefont {Alloul}},\ }\href
  {\doibase 10.1103/PhysRevLett.103.057001} {\bibfield  {journal} {\bibinfo
  {journal} {Phys. Rev. Lett.}\ }\textbf {\bibinfo {volume} {103}},\ \bibinfo
  {pages} {057001} (\bibinfo {year} {2009})}\BibitemShut {NoStop}%
\bibitem [{\citenamefont {Lee}\ \emph {et~al.}(2009)\citenamefont {Lee},
  \citenamefont {Bartkowiak}, \citenamefont {Park}, \citenamefont {Lee},
  \citenamefont {Kim}, \citenamefont {Sung}, \citenamefont {Cho}, \citenamefont
  {Jung}, \citenamefont {Kim},\ and\ \citenamefont {Lee}}]{LeeBartkowiak2009}%
  \BibitemOpen
  \bibfield  {author} {\bibinfo {author} {\bibfnamefont {H.-S.}\ \bibnamefont
  {Lee}}, \bibinfo {author} {\bibfnamefont {M.}~\bibnamefont {Bartkowiak}},
  \bibinfo {author} {\bibfnamefont {J.-H.}\ \bibnamefont {Park}}, \bibinfo
  {author} {\bibfnamefont {J.-Y.}\ \bibnamefont {Lee}}, \bibinfo {author}
  {\bibfnamefont {J.-Y.}\ \bibnamefont {Kim}}, \bibinfo {author} {\bibfnamefont
  {N.-H.}\ \bibnamefont {Sung}}, \bibinfo {author} {\bibfnamefont {B.~K.}\
  \bibnamefont {Cho}}, \bibinfo {author} {\bibfnamefont {C.-U.}\ \bibnamefont
  {Jung}}, \bibinfo {author} {\bibfnamefont {J.~S.}\ \bibnamefont {Kim}}, \
  and\ \bibinfo {author} {\bibfnamefont {H.-J.}\ \bibnamefont {Lee}},\ }\href
  {\doibase 10.1103/PhysRevB.80.144512} {\bibfield  {journal} {\bibinfo
  {journal} {Phys. Rev. B}\ }\textbf {\bibinfo {volume} {80}},\ \bibinfo
  {pages} {144512} (\bibinfo {year} {2009})}\BibitemShut {NoStop}%
\bibitem [{\citenamefont {Demsar}\ \emph {et~al.}(1999)\citenamefont {Demsar},
  \citenamefont {Podobnik}, \citenamefont {Kabanov}, \citenamefont {Wolf},\
  and\ \citenamefont {Mihailovic}}]{DemsarPodobnik1999}%
  \BibitemOpen
  \bibfield  {author} {\bibinfo {author} {\bibfnamefont {J.}~\bibnamefont
  {Demsar}}, \bibinfo {author} {\bibfnamefont {B.}~\bibnamefont {Podobnik}},
  \bibinfo {author} {\bibfnamefont {V.~V.}\ \bibnamefont {Kabanov}}, \bibinfo
  {author} {\bibfnamefont {T.}~\bibnamefont {Wolf}}, \ and\ \bibinfo {author}
  {\bibfnamefont {D.}~\bibnamefont {Mihailovic}},\ }\href {\doibase
  10.1103/PhysRevLett.82.4918} {\bibfield  {journal} {\bibinfo  {journal}
  {Phys. Rev. Lett.}\ }\textbf {\bibinfo {volume} {82}},\ \bibinfo {pages}
  {4918} (\bibinfo {year} {1999})}\BibitemShut {NoStop}%
\bibitem [{\citenamefont {Kaindl}\ \emph {et~al.}(2000)\citenamefont {Kaindl},
  \citenamefont {Woerner}, \citenamefont {Elsaesser}, \citenamefont {Smith},
  \citenamefont {Ryan}, \citenamefont {Farnan}, \citenamefont {McCurry},\ and\
  \citenamefont {Walmsley}}]{KaindlWoerner2000}%
  \BibitemOpen
  \bibfield  {author} {\bibinfo {author} {\bibfnamefont {R.}~\bibnamefont
  {Kaindl}}, \bibinfo {author} {\bibfnamefont {M.}~\bibnamefont {Woerner}},
  \bibinfo {author} {\bibfnamefont {T.}~\bibnamefont {Elsaesser}}, \bibinfo
  {author} {\bibfnamefont {D.}~\bibnamefont {Smith}}, \bibinfo {author}
  {\bibfnamefont {J.}~\bibnamefont {Ryan}}, \bibinfo {author} {\bibfnamefont
  {G.}~\bibnamefont {Farnan}}, \bibinfo {author} {\bibfnamefont
  {M.}~\bibnamefont {McCurry}}, \ and\ \bibinfo {author} {\bibfnamefont
  {D.}~\bibnamefont {Walmsley}},\ }\href@noop {} {\bibfield  {journal}
  {\bibinfo  {journal} {Science}\ }\textbf {\bibinfo {volume} {287}},\ \bibinfo
  {pages} {470} (\bibinfo {year} {2000})}\BibitemShut {NoStop}%
\bibitem [{\citenamefont {Averitt}\ \emph {et~al.}(2001)\citenamefont
  {Averitt}, \citenamefont {Rodriguez}, \citenamefont {Lobad}, \citenamefont
  {Siders}, \citenamefont {Trugman},\ and\ \citenamefont
  {Taylor}}]{AverittRodriguez2001}%
  \BibitemOpen
  \bibfield  {author} {\bibinfo {author} {\bibfnamefont {R.~D.}\ \bibnamefont
  {Averitt}}, \bibinfo {author} {\bibfnamefont {G.}~\bibnamefont {Rodriguez}},
  \bibinfo {author} {\bibfnamefont {A.~I.}\ \bibnamefont {Lobad}}, \bibinfo
  {author} {\bibfnamefont {J.~L.~W.}\ \bibnamefont {Siders}}, \bibinfo {author}
  {\bibfnamefont {S.~A.}\ \bibnamefont {Trugman}}, \ and\ \bibinfo {author}
  {\bibfnamefont {A.~J.}\ \bibnamefont {Taylor}},\ }\href {\doibase
  10.1103/PhysRevB.63.140502} {\bibfield  {journal} {\bibinfo  {journal} {Phys.
  Rev. B}\ }\textbf {\bibinfo {volume} {63}},\ \bibinfo {pages} {140502}
  (\bibinfo {year} {2001})}\BibitemShut {NoStop}%
\bibitem [{\citenamefont {Segre}\ \emph {et~al.}(2002)\citenamefont {Segre},
  \citenamefont {Gedik}, \citenamefont {Orenstein}, \citenamefont {Bonn},
  \citenamefont {Liang},\ and\ \citenamefont {Hardy}}]{SegreGedik2002}%
  \BibitemOpen
  \bibfield  {author} {\bibinfo {author} {\bibfnamefont {G.~P.}\ \bibnamefont
  {Segre}}, \bibinfo {author} {\bibfnamefont {N.}~\bibnamefont {Gedik}},
  \bibinfo {author} {\bibfnamefont {J.}~\bibnamefont {Orenstein}}, \bibinfo
  {author} {\bibfnamefont {D.~A.}\ \bibnamefont {Bonn}}, \bibinfo {author}
  {\bibfnamefont {R.}~\bibnamefont {Liang}}, \ and\ \bibinfo {author}
  {\bibfnamefont {W.~N.}\ \bibnamefont {Hardy}},\ }\href {\doibase
  10.1103/PhysRevLett.88.137001} {\bibfield  {journal} {\bibinfo  {journal}
  {Phys. Rev. Lett.}\ }\textbf {\bibinfo {volume} {88}},\ \bibinfo {pages}
  {137001} (\bibinfo {year} {2002})}\BibitemShut {NoStop}%
\bibitem [{\citenamefont {Kusar}\ \emph {et~al.}(2005)\citenamefont {Kusar},
  \citenamefont {Demsar}, \citenamefont {Mihailovic},\ and\ \citenamefont
  {Sugai}}]{KusarDemsar2005}%
  \BibitemOpen
  \bibfield  {author} {\bibinfo {author} {\bibfnamefont {P.}~\bibnamefont
  {Kusar}}, \bibinfo {author} {\bibfnamefont {J.}~\bibnamefont {Demsar}},
  \bibinfo {author} {\bibfnamefont {D.}~\bibnamefont {Mihailovic}}, \ and\
  \bibinfo {author} {\bibfnamefont {S.}~\bibnamefont {Sugai}},\ }\href
  {\doibase 10.1103/PhysRevB.72.014544} {\bibfield  {journal} {\bibinfo
  {journal} {Phys. Rev. B}\ }\textbf {\bibinfo {volume} {72}},\ \bibinfo
  {pages} {014544} (\bibinfo {year} {2005})}\BibitemShut {NoStop}%
\bibitem [{\citenamefont {Liu}\ \emph {et~al.}(2008{\natexlab{b}})\citenamefont
  {Liu}, \citenamefont {Toda}, \citenamefont {Shimatake}, \citenamefont
  {Momono}, \citenamefont {Oda},\ and\ \citenamefont {Ido}}]{LiuToda2008}%
  \BibitemOpen
  \bibfield  {author} {\bibinfo {author} {\bibfnamefont {Y.~H.}\ \bibnamefont
  {Liu}}, \bibinfo {author} {\bibfnamefont {Y.}~\bibnamefont {Toda}}, \bibinfo
  {author} {\bibfnamefont {K.}~\bibnamefont {Shimatake}}, \bibinfo {author}
  {\bibfnamefont {N.}~\bibnamefont {Momono}}, \bibinfo {author} {\bibfnamefont
  {M.}~\bibnamefont {Oda}}, \ and\ \bibinfo {author} {\bibfnamefont
  {M.}~\bibnamefont {Ido}},\ }\href {\doibase 10.1103/PhysRevLett.101.137003}
  {\bibfield  {journal} {\bibinfo  {journal} {Phys. Rev. Lett.}\ }\textbf
  {\bibinfo {volume} {101}},\ \bibinfo {pages} {137003} (\bibinfo {year}
  {2008}{\natexlab{b}})}\BibitemShut {NoStop}%
\bibitem [{\citenamefont {Ning}\ \emph {et~al.}(2008)\citenamefont {Ning},
  \citenamefont {Yan-Feng}, \citenamefont {Ji-Min}, \citenamefont {Shi-Ping},
  \citenamefont {Qian-Sheng}, \citenamefont {Zhi-Guo},\ and\ \citenamefont
  {Pan-Ming}}]{CaoWei2008}%
  \BibitemOpen
  \bibfield  {author} {\bibinfo {author} {\bibfnamefont {C.}~\bibnamefont
  {Ning}}, \bibinfo {author} {\bibfnamefont {W.}~\bibnamefont {Yan-Feng}},
  \bibinfo {author} {\bibfnamefont {Z.}~\bibnamefont {Ji-Min}}, \bibinfo
  {author} {\bibfnamefont {Z.}~\bibnamefont {Shi-Ping}}, \bibinfo {author}
  {\bibfnamefont {Y.}~\bibnamefont {Qian-Sheng}}, \bibinfo {author}
  {\bibfnamefont {Z.}~\bibnamefont {Zhi-Guo}}, \ and\ \bibinfo {author}
  {\bibfnamefont {F.}~\bibnamefont {Pan-Ming}},\ }\href
  {http://stacks.iop.org/0256-307X/25/i=6/a=092} {\bibfield  {journal}
  {\bibinfo  {journal} {Chinese Physics Letters}\ }\textbf {\bibinfo {volume}
  {25}},\ \bibinfo {pages} {2257} (\bibinfo {year} {2008})}\BibitemShut
  {NoStop}%
\bibitem [{\citenamefont {Mertelj}\ \emph
  {et~al.}(2009{\natexlab{b}})\citenamefont {Mertelj}, \citenamefont {Kabanov},
  \citenamefont {Gadermaier}, \citenamefont {Zhigadlo}, \citenamefont
  {Katrych}, \citenamefont {Bukowski}, \citenamefont {Karpinski},\ and\
  \citenamefont {Mihailovic}}]{MerteljKabanov2009jsnm}%
  \BibitemOpen
  \bibfield  {author} {\bibinfo {author} {\bibfnamefont {T.}~\bibnamefont
  {Mertelj}}, \bibinfo {author} {\bibfnamefont {V.}~\bibnamefont {Kabanov}},
  \bibinfo {author} {\bibfnamefont {C.}~\bibnamefont {Gadermaier}}, \bibinfo
  {author} {\bibfnamefont {N.}~\bibnamefont {Zhigadlo}}, \bibinfo {author}
  {\bibfnamefont {S.}~\bibnamefont {Katrych}}, \bibinfo {author} {\bibfnamefont
  {Z.}~\bibnamefont {Bukowski}}, \bibinfo {author} {\bibfnamefont
  {J.}~\bibnamefont {Karpinski}}, \ and\ \bibinfo {author} {\bibfnamefont
  {D.}~\bibnamefont {Mihailovic}},\ }\href@noop {} {\bibfield  {journal}
  {\bibinfo  {journal} {Journal of Superconductivity and Novel Magnetism}\
  }\textbf {\bibinfo {volume} {22}},\ \bibinfo {pages} {575} (\bibinfo {year}
  {2009}{\natexlab{b}})}\BibitemShut {NoStop}%
\bibitem [{\citenamefont {Mertelj}\ \emph {et~al.}(2010)\citenamefont
  {Mertelj}, \citenamefont {Kusar}, \citenamefont {Kabanov}, \citenamefont
  {Stojchevska}, \citenamefont {Zhigadlo}, \citenamefont {Katrych},
  \citenamefont {Bukowski}, \citenamefont {Karpinski}, \citenamefont
  {Weyeneth},\ and\ \citenamefont {Mihailovic}}]{MerteljKusar2010}%
  \BibitemOpen
  \bibfield  {author} {\bibinfo {author} {\bibfnamefont {T.}~\bibnamefont
  {Mertelj}}, \bibinfo {author} {\bibfnamefont {P.}~\bibnamefont {Kusar}},
  \bibinfo {author} {\bibfnamefont {V.~V.}\ \bibnamefont {Kabanov}}, \bibinfo
  {author} {\bibfnamefont {L.}~\bibnamefont {Stojchevska}}, \bibinfo {author}
  {\bibfnamefont {N.~D.}\ \bibnamefont {Zhigadlo}}, \bibinfo {author}
  {\bibfnamefont {S.}~\bibnamefont {Katrych}}, \bibinfo {author} {\bibfnamefont
  {Z.}~\bibnamefont {Bukowski}}, \bibinfo {author} {\bibfnamefont
  {J.}~\bibnamefont {Karpinski}}, \bibinfo {author} {\bibfnamefont
  {S.}~\bibnamefont {Weyeneth}}, \ and\ \bibinfo {author} {\bibfnamefont
  {D.}~\bibnamefont {Mihailovic}},\ }\href {\doibase
  10.1103/PhysRevB.81.224504} {\bibfield  {journal} {\bibinfo  {journal} {Phys.
  Rev. B}\ }\textbf {\bibinfo {volume} {81}},\ \bibinfo {pages} {224504}
  (\bibinfo {year} {2010})}\BibitemShut {NoStop}%
\bibitem [{\citenamefont {Torchinsky}\ \emph {et~al.}(2010)\citenamefont
  {Torchinsky}, \citenamefont {Chen}, \citenamefont {Luo}, \citenamefont
  {Wang},\ and\ \citenamefont {Gedik}}]{TorchinskyChen2010}%
  \BibitemOpen
  \bibfield  {author} {\bibinfo {author} {\bibfnamefont {D.~H.}\ \bibnamefont
  {Torchinsky}}, \bibinfo {author} {\bibfnamefont {G.~F.}\ \bibnamefont
  {Chen}}, \bibinfo {author} {\bibfnamefont {J.~L.}\ \bibnamefont {Luo}},
  \bibinfo {author} {\bibfnamefont {N.~L.}\ \bibnamefont {Wang}}, \ and\
  \bibinfo {author} {\bibfnamefont {N.}~\bibnamefont {Gedik}},\ }\href
  {\doibase 10.1103/PhysRevLett.105.027005} {\bibfield  {journal} {\bibinfo
  {journal} {Phys. Rev. Lett.}\ }\textbf {\bibinfo {volume} {105}},\ \bibinfo
  {pages} {027005} (\bibinfo {year} {2010})}\BibitemShut {NoStop}%
\bibitem [{\citenamefont {Chia}\ \emph {et~al.}(2010)\citenamefont {Chia},
  \citenamefont {Talbayev}, \citenamefont {Zhu}, \citenamefont {Yuan},
  \citenamefont {Park}, \citenamefont {Thompson}, \citenamefont {Panagopoulos},
  \citenamefont {Chen}, \citenamefont {Luo}, \citenamefont {Wang},\ and\
  \citenamefont {Taylor}}]{ChiaTalbayev2010}%
  \BibitemOpen
  \bibfield  {author} {\bibinfo {author} {\bibfnamefont {E.~E.~M.}\
  \bibnamefont {Chia}}, \bibinfo {author} {\bibfnamefont {D.}~\bibnamefont
  {Talbayev}}, \bibinfo {author} {\bibfnamefont {J.-X.}\ \bibnamefont {Zhu}},
  \bibinfo {author} {\bibfnamefont {H.~Q.}\ \bibnamefont {Yuan}}, \bibinfo
  {author} {\bibfnamefont {T.}~\bibnamefont {Park}}, \bibinfo {author}
  {\bibfnamefont {J.~D.}\ \bibnamefont {Thompson}}, \bibinfo {author}
  {\bibfnamefont {C.}~\bibnamefont {Panagopoulos}}, \bibinfo {author}
  {\bibfnamefont {G.~F.}\ \bibnamefont {Chen}}, \bibinfo {author}
  {\bibfnamefont {J.~L.}\ \bibnamefont {Luo}}, \bibinfo {author} {\bibfnamefont
  {N.~L.}\ \bibnamefont {Wang}}, \ and\ \bibinfo {author} {\bibfnamefont
  {A.~J.}\ \bibnamefont {Taylor}},\ }\href {\doibase
  10.1103/PhysRevLett.104.027003} {\bibfield  {journal} {\bibinfo  {journal}
  {Phys. Rev. Lett.}\ }\textbf {\bibinfo {volume} {104}},\ \bibinfo {pages}
  {027003} (\bibinfo {year} {2010})}\BibitemShut {NoStop}%
\bibitem [{\citenamefont {Stojchevska}\ \emph {et~al.}(2010)\citenamefont
  {Stojchevska}, \citenamefont {Kusar}, \citenamefont {Mertelj}, \citenamefont
  {Kabanov}, \citenamefont {Lin}, \citenamefont {Cao}, \citenamefont {Xu},\
  and\ \citenamefont {Mihailovic}}]{StojchevskaKusar2010}%
  \BibitemOpen
  \bibfield  {author} {\bibinfo {author} {\bibfnamefont {L.}~\bibnamefont
  {Stojchevska}}, \bibinfo {author} {\bibfnamefont {P.}~\bibnamefont {Kusar}},
  \bibinfo {author} {\bibfnamefont {T.}~\bibnamefont {Mertelj}}, \bibinfo
  {author} {\bibfnamefont {V.~V.}\ \bibnamefont {Kabanov}}, \bibinfo {author}
  {\bibfnamefont {X.}~\bibnamefont {Lin}}, \bibinfo {author} {\bibfnamefont
  {G.~H.}\ \bibnamefont {Cao}}, \bibinfo {author} {\bibfnamefont {Z.~A.}\
  \bibnamefont {Xu}}, \ and\ \bibinfo {author} {\bibfnamefont {D.}~\bibnamefont
  {Mihailovic}},\ }\href {\doibase 10.1103/PhysRevB.82.012505} {\bibfield
  {journal} {\bibinfo  {journal} {Phys. Rev. B}\ }\textbf {\bibinfo {volume}
  {82}},\ \bibinfo {pages} {012505} (\bibinfo {year} {2010})}\BibitemShut
  {NoStop}%
\bibitem [{\citenamefont {Gong}\ \emph {et~al.}(2010)\citenamefont {Gong},
  \citenamefont {Lai}, \citenamefont {Nosach}, \citenamefont {Li},
  \citenamefont {Cao}, \citenamefont {Xu},\ and\ \citenamefont
  {Ren}}]{GongLai2010}%
  \BibitemOpen
  \bibfield  {author} {\bibinfo {author} {\bibfnamefont {Y.}~\bibnamefont
  {Gong}}, \bibinfo {author} {\bibfnamefont {W.}~\bibnamefont {Lai}}, \bibinfo
  {author} {\bibfnamefont {T.}~\bibnamefont {Nosach}}, \bibinfo {author}
  {\bibfnamefont {L.~J.}\ \bibnamefont {Li}}, \bibinfo {author} {\bibfnamefont
  {G.~H.}\ \bibnamefont {Cao}}, \bibinfo {author} {\bibfnamefont {Z.~A.}\
  \bibnamefont {Xu}}, \ and\ \bibinfo {author} {\bibfnamefont {Y.~H.}\
  \bibnamefont {Ren}},\ }\href
  {http://stacks.iop.org/1367-2630/12/i=12/a=123003} {\bibfield  {journal}
  {\bibinfo  {journal} {New Journal of Physics}\ }\textbf {\bibinfo {volume}
  {12}},\ \bibinfo {pages} {123003} (\bibinfo {year} {2010})}\BibitemShut
  {NoStop}%
\bibitem [{\citenamefont {Mansart}\ \emph {et~al.}(2010)\citenamefont
  {Mansart}, \citenamefont {Boschetto}, \citenamefont {Savoia}, \citenamefont
  {Rullier-Albenque}, \citenamefont {Bouquet}, \citenamefont {Papalazarou},
  \citenamefont {Forget}, \citenamefont {Colson}, \citenamefont {Rousse},\ and\
  \citenamefont {Marsi}}]{MansartBoschetto2010}%
  \BibitemOpen
  \bibfield  {author} {\bibinfo {author} {\bibfnamefont {B.}~\bibnamefont
  {Mansart}}, \bibinfo {author} {\bibfnamefont {D.}~\bibnamefont {Boschetto}},
  \bibinfo {author} {\bibfnamefont {A.}~\bibnamefont {Savoia}}, \bibinfo
  {author} {\bibfnamefont {F.}~\bibnamefont {Rullier-Albenque}}, \bibinfo
  {author} {\bibfnamefont {F.}~\bibnamefont {Bouquet}}, \bibinfo {author}
  {\bibfnamefont {E.}~\bibnamefont {Papalazarou}}, \bibinfo {author}
  {\bibfnamefont {A.}~\bibnamefont {Forget}}, \bibinfo {author} {\bibfnamefont
  {D.}~\bibnamefont {Colson}}, \bibinfo {author} {\bibfnamefont
  {A.}~\bibnamefont {Rousse}}, \ and\ \bibinfo {author} {\bibfnamefont
  {M.}~\bibnamefont {Marsi}},\ }\href {\doibase 10.1103/PhysRevB.82.024513}
  {\bibfield  {journal} {\bibinfo  {journal} {Phys. Rev. B}\ }\textbf {\bibinfo
  {volume} {82}},\ \bibinfo {pages} {024513} (\bibinfo {year}
  {2010})}\BibitemShut {NoStop}%
\bibitem [{\citenamefont {Torchinsky}\ \emph {et~al.}(2011)\citenamefont
  {Torchinsky}, \citenamefont {McIver}, \citenamefont {Hsieh}, \citenamefont
  {Chen}, \citenamefont {Luo}, \citenamefont {Wang},\ and\ \citenamefont
  {Gedik}}]{TorchinskyMcIver2011}%
  \BibitemOpen
  \bibfield  {author} {\bibinfo {author} {\bibfnamefont {D.~H.}\ \bibnamefont
  {Torchinsky}}, \bibinfo {author} {\bibfnamefont {J.~W.}\ \bibnamefont
  {McIver}}, \bibinfo {author} {\bibfnamefont {D.}~\bibnamefont {Hsieh}},
  \bibinfo {author} {\bibfnamefont {G.~F.}\ \bibnamefont {Chen}}, \bibinfo
  {author} {\bibfnamefont {J.~L.}\ \bibnamefont {Luo}}, \bibinfo {author}
  {\bibfnamefont {N.~L.}\ \bibnamefont {Wang}}, \ and\ \bibinfo {author}
  {\bibfnamefont {N.}~\bibnamefont {Gedik}},\ }\href {\doibase
  10.1103/PhysRevB.84.104518} {\bibfield  {journal} {\bibinfo  {journal} {Phys.
  Rev. B}\ }\textbf {\bibinfo {volume} {84}},\ \bibinfo {pages} {104518}
  (\bibinfo {year} {2011})}\BibitemShut {NoStop}%
\bibitem [{\citenamefont {Chu}\ \emph {et~al.}(2009)\citenamefont {Chu},
  \citenamefont {Analytis}, \citenamefont {Kucharczyk},\ and\ \citenamefont
  {Fisher}}]{ChuAnalytis2009}%
  \BibitemOpen
  \bibfield  {author} {\bibinfo {author} {\bibfnamefont {J.-H.}\ \bibnamefont
  {Chu}}, \bibinfo {author} {\bibfnamefont {J.~G.}\ \bibnamefont {Analytis}},
  \bibinfo {author} {\bibfnamefont {C.}~\bibnamefont {Kucharczyk}}, \ and\
  \bibinfo {author} {\bibfnamefont {I.~R.}\ \bibnamefont {Fisher}},\ }\href
  {\doibase 10.1103/PhysRevB.79.014506} {\bibfield  {journal} {\bibinfo
  {journal} {Phys. Rev. B}\ }\textbf {\bibinfo {volume} {79}},\ \bibinfo
  {pages} {014506} (\bibinfo {year} {2009})}\BibitemShut {NoStop}%
\bibitem [{\citenamefont {Lester}\ \emph {et~al.}(2009)\citenamefont {Lester},
  \citenamefont {Chu}, \citenamefont {Analytis}, \citenamefont {Capelli},
  \citenamefont {Erickson}, \citenamefont {Condron}, \citenamefont {Toney},
  \citenamefont {Fisher},\ and\ \citenamefont {Hayden}}]{LesterChu2009}%
  \BibitemOpen
  \bibfield  {author} {\bibinfo {author} {\bibfnamefont {C.}~\bibnamefont
  {Lester}}, \bibinfo {author} {\bibfnamefont {J.-H.}\ \bibnamefont {Chu}},
  \bibinfo {author} {\bibfnamefont {J.~G.}\ \bibnamefont {Analytis}}, \bibinfo
  {author} {\bibfnamefont {S.~C.}\ \bibnamefont {Capelli}}, \bibinfo {author}
  {\bibfnamefont {A.~S.}\ \bibnamefont {Erickson}}, \bibinfo {author}
  {\bibfnamefont {C.~L.}\ \bibnamefont {Condron}}, \bibinfo {author}
  {\bibfnamefont {M.~F.}\ \bibnamefont {Toney}}, \bibinfo {author}
  {\bibfnamefont {I.~R.}\ \bibnamefont {Fisher}}, \ and\ \bibinfo {author}
  {\bibfnamefont {S.~M.}\ \bibnamefont {Hayden}},\ }\href {\doibase
  10.1103/PhysRevB.79.144523} {\bibfield  {journal} {\bibinfo  {journal} {Phys.
  Rev. B}\ }\textbf {\bibinfo {volume} {79}},\ \bibinfo {pages} {144523}
  (\bibinfo {year} {2009})}\BibitemShut {NoStop}%
\bibitem [{Note1()}]{Note1}%
  \BibitemOpen
  \bibinfo {note} {We observed no pump polarization dependence of the response
  at fixed probe polarization.}\BibitemShut {Stop}%
\bibitem [{\citenamefont {Kusar}\ \emph {et~al.}(2008)\citenamefont {Kusar},
  \citenamefont {Kabanov}, \citenamefont {Demsar}, \citenamefont {Mertelj},
  \citenamefont {Sugai},\ and\ \citenamefont {Mihailovic}}]{KusarKabanov2008}%
  \BibitemOpen
  \bibfield  {author} {\bibinfo {author} {\bibfnamefont {P.}~\bibnamefont
  {Kusar}}, \bibinfo {author} {\bibfnamefont {V.}~\bibnamefont {Kabanov}},
  \bibinfo {author} {\bibfnamefont {J.}~\bibnamefont {Demsar}}, \bibinfo
  {author} {\bibfnamefont {T.}~\bibnamefont {Mertelj}}, \bibinfo {author}
  {\bibfnamefont {S.}~\bibnamefont {Sugai}}, \ and\ \bibinfo {author}
  {\bibfnamefont {D.}~\bibnamefont {Mihailovic}},\ }\href@noop {} {\bibfield
  {journal} {\bibinfo  {journal} {Physical Review Letters}\ }\textbf {\bibinfo
  {volume} {101}},\ \bibinfo {pages} {227001} (\bibinfo {year}
  {2008})}\BibitemShut {NoStop}%
\bibitem [{Note2()}]{Note2}%
  \BibitemOpen
  \bibinfo {note} {The anisotropy indicates a preferential ordering of the
  orthorhombic twin domains in the probed volume due to the anisotropic surface
  strain.}\BibitemShut {Stop}%
\bibitem [{Note3()}]{Note3}%
  \BibitemOpen
  \bibinfo {note} {The critical temperature was determined from our optical
  measurements based on the sample holder temperature and is apparently lower
  than the phase diagram {[}Fig. \ref {fig:DR-2D} (e){]} value due to the
  sample heating by the laser.}\BibitemShut {Stop}%
\bibitem [{\citenamefont {Bari\ifmmode \check{s}\else
  \v{s}\fi{}i\ifmmode~\acute{c}\else \'{c}\fi{}}\ \emph
  {et~al.}(2010)\citenamefont {Bari\ifmmode \check{s}\else
  \v{s}\fi{}i\ifmmode~\acute{c}\else \'{c}\fi{}}, \citenamefont {Wu},
  \citenamefont {Dressel}, \citenamefont {Li}, \citenamefont {Cao},\ and\
  \citenamefont {Xu}}]{BarisicWu2010}%
  \BibitemOpen
  \bibfield  {author} {\bibinfo {author} {\bibfnamefont {N.}~\bibnamefont
  {Bari\ifmmode \check{s}\else \v{s}\fi{}i\ifmmode~\acute{c}\else \'{c}\fi{}}},
  \bibinfo {author} {\bibfnamefont {D.}~\bibnamefont {Wu}}, \bibinfo {author}
  {\bibfnamefont {M.}~\bibnamefont {Dressel}}, \bibinfo {author} {\bibfnamefont
  {L.~J.}\ \bibnamefont {Li}}, \bibinfo {author} {\bibfnamefont {G.~H.}\
  \bibnamefont {Cao}}, \ and\ \bibinfo {author} {\bibfnamefont {Z.~A.}\
  \bibnamefont {Xu}},\ }\href {\doibase 10.1103/PhysRevB.82.054518} {\bibfield
  {journal} {\bibinfo  {journal} {Phys. Rev. B}\ }\textbf {\bibinfo {volume}
  {82}},\ \bibinfo {pages} {054518} (\bibinfo {year} {2010})}\BibitemShut
  {NoStop}%
\bibitem [{Note4()}]{Note4}%
  \BibitemOpen
  \bibinfo {note} {We calculated $U_{\protect \mathrm {c}}$ from the heat
  capacity data in Ref. \protect \rev@citealpnum
  {HardyBurger2010}.}\BibitemShut {Stop}%
\bibitem [{\citenamefont {Stojchevska}\ \emph {et~al.}(2011)\citenamefont
  {Stojchevska}, \citenamefont {Kusar}, \citenamefont {Mertelj}, \citenamefont
  {Kabanov}, \citenamefont {Toda}, \citenamefont {Yao},\ and\ \citenamefont
  {Mihailovic}}]{StojchevskaKusar2011}%
  \BibitemOpen
  \bibfield  {author} {\bibinfo {author} {\bibfnamefont {L.}~\bibnamefont
  {Stojchevska}}, \bibinfo {author} {\bibfnamefont {P.}~\bibnamefont {Kusar}},
  \bibinfo {author} {\bibfnamefont {T.}~\bibnamefont {Mertelj}}, \bibinfo
  {author} {\bibfnamefont {V.~V.}\ \bibnamefont {Kabanov}}, \bibinfo {author}
  {\bibfnamefont {Y.}~\bibnamefont {Toda}}, \bibinfo {author} {\bibfnamefont
  {X.}~\bibnamefont {Yao}}, \ and\ \bibinfo {author} {\bibfnamefont
  {D.}~\bibnamefont {Mihailovic}},\ }\href {\doibase
  10.1103/PhysRevB.84.180507} {\bibfield  {journal} {\bibinfo  {journal} {Phys.
  Rev. B}\ }\textbf {\bibinfo {volume} {84}},\ \bibinfo {pages} {180507}
  (\bibinfo {year} {2011})}\BibitemShut {NoStop}%
\bibitem [{Note5()}]{Note5}%
  \BibitemOpen
  \bibinfo {note} {The weak non monotonic temporal dependence of the SC
  component during the plateau could not be reliably identified as an intrinsic
  effect and is attributed to a weak $T$-dependence of the subtracted normal
  state response.}\BibitemShut {Stop}%
\bibitem [{\citenamefont {Mattis}\ and\ \citenamefont
  {Bardeen}(1958)}]{MattisBardeen1985}%
  \BibitemOpen
  \bibfield  {author} {\bibinfo {author} {\bibfnamefont {D.~C.}\ \bibnamefont
  {Mattis}}\ and\ \bibinfo {author} {\bibfnamefont {J.}~\bibnamefont
  {Bardeen}},\ }\href {\doibase 10.1103/PhysRev.111.412} {\bibfield  {journal}
  {\bibinfo  {journal} {Phys. Rev.}\ }\textbf {\bibinfo {volume} {111}},\
  \bibinfo {pages} {412} (\bibinfo {year} {1958})}\BibitemShut {NoStop}%
\bibitem [{\citenamefont {Hardy}\ \emph {et~al.}(2010)\citenamefont {Hardy},
  \citenamefont {Burger}, \citenamefont {Wolf}, \citenamefont {Fisher},
  \citenamefont {Schweiss}, \citenamefont {Adelmann}, \citenamefont {Heid},
  \citenamefont {Fromknecht}, \citenamefont {Eder}, \citenamefont {Ernst},
  \citenamefont {v.~Löhneysen},\ and\ \citenamefont
  {Meingast}}]{HardyBurger2010}%
  \BibitemOpen
  \bibfield  {author} {\bibinfo {author} {\bibfnamefont {F.}~\bibnamefont
  {Hardy}}, \bibinfo {author} {\bibfnamefont {P.}~\bibnamefont {Burger}},
  \bibinfo {author} {\bibfnamefont {T.}~\bibnamefont {Wolf}}, \bibinfo {author}
  {\bibfnamefont {R.~A.}\ \bibnamefont {Fisher}}, \bibinfo {author}
  {\bibfnamefont {P.}~\bibnamefont {Schweiss}}, \bibinfo {author}
  {\bibfnamefont {P.}~\bibnamefont {Adelmann}}, \bibinfo {author}
  {\bibfnamefont {R.}~\bibnamefont {Heid}}, \bibinfo {author} {\bibfnamefont
  {R.}~\bibnamefont {Fromknecht}}, \bibinfo {author} {\bibfnamefont
  {R.}~\bibnamefont {Eder}}, \bibinfo {author} {\bibfnamefont {D.}~\bibnamefont
  {Ernst}}, \bibinfo {author} {\bibfnamefont {H.}~\bibnamefont {v.~Löhneysen}},
  \ and\ \bibinfo {author} {\bibfnamefont {C.}~\bibnamefont {Meingast}},\
  }\href {http://stacks.iop.org/0295-5075/91/i=4/a=47008} {\bibfield  {journal}
  {\bibinfo  {journal} {EPL (Europhysics Letters)}\ }\textbf {\bibinfo {volume}
  {91}},\ \bibinfo {pages} {47008} (\bibinfo {year} {2010})}\BibitemShut
  {NoStop}%
\bibitem [{\citenamefont {Tropeano}\ \emph {et~al.}(2008)\citenamefont
  {Tropeano}, \citenamefont {Martinelli}, \citenamefont {Palenzona},
  \citenamefont {Bellingeri}, \citenamefont {d'Agliano}, \citenamefont
  {Nguyen}, \citenamefont {Affronte},\ and\ \citenamefont
  {Putti}}]{TropeanoMartinelli2008}%
  \BibitemOpen
  \bibfield  {author} {\bibinfo {author} {\bibfnamefont {M.}~\bibnamefont
  {Tropeano}}, \bibinfo {author} {\bibfnamefont {A.}~\bibnamefont
  {Martinelli}}, \bibinfo {author} {\bibfnamefont {A.}~\bibnamefont
  {Palenzona}}, \bibinfo {author} {\bibfnamefont {E.}~\bibnamefont
  {Bellingeri}}, \bibinfo {author} {\bibfnamefont {E.~G.}\ \bibnamefont
  {d'Agliano}}, \bibinfo {author} {\bibfnamefont {T.~D.}\ \bibnamefont
  {Nguyen}}, \bibinfo {author} {\bibfnamefont {M.}~\bibnamefont {Affronte}}, \
  and\ \bibinfo {author} {\bibfnamefont {M.}~\bibnamefont {Putti}},\ }\href
  {\doibase 10.1103/PhysRevB.78.094518} {\bibfield  {journal} {\bibinfo
  {journal} {Physical Review B (Condensed Matter and Materials Physics)}\
  }\textbf {\bibinfo {volume} {78}},\ \bibinfo {eid} {094518} (\bibinfo {year}
  {2008})}\BibitemShut {NoStop}%
\bibitem [{\citenamefont {Rettig}\ \emph {et~al.}(2010)\citenamefont {Rettig},
  \citenamefont {Cort{\~A}{\v{S}}s}, \citenamefont {Thirupathaiah},
  \citenamefont {Gegenwart}, \citenamefont {Jeevan}, \citenamefont {Wolf},
  \citenamefont {Bovensiepen}, \citenamefont {Wolf}, \citenamefont
  {D{\~A}trr},\ and\ \citenamefont {Fink}}]{RettigCortes2010}%
  \BibitemOpen
  \bibfield  {author} {\bibinfo {author} {\bibfnamefont {L.}~\bibnamefont
  {Rettig}}, \bibinfo {author} {\bibfnamefont {R.}~\bibnamefont
  {Cort{\~A}{\v{S}}s}}, \bibinfo {author} {\bibfnamefont {S.}~\bibnamefont
  {Thirupathaiah}}, \bibinfo {author} {\bibfnamefont {P.}~\bibnamefont
  {Gegenwart}}, \bibinfo {author} {\bibfnamefont {H.}~\bibnamefont {Jeevan}},
  \bibinfo {author} {\bibfnamefont {T.}~\bibnamefont {Wolf}}, \bibinfo {author}
  {\bibfnamefont {U.}~\bibnamefont {Bovensiepen}}, \bibinfo {author}
  {\bibfnamefont {M.}~\bibnamefont {Wolf}}, \bibinfo {author} {\bibfnamefont
  {H.}~\bibnamefont {D{\~A}trr}}, \ and\ \bibinfo {author} {\bibfnamefont
  {J.}~\bibnamefont {Fink}},\ }\href@noop {} {\bibfield  {journal} {\bibinfo
  {journal} {Arxiv preprint arXiv:1008.1561}\ } (\bibinfo {year}
  {2010})}\BibitemShut {NoStop}%
\bibitem [{Note6()}]{Note6}%
  \BibitemOpen
  \bibinfo {note} {Due to the presence of the surface-strain bias the
  transition is, strictly speaking, a crossover.\cite
  {ChuAnalytis2010}}\BibitemShut {NoStop}%
\bibitem [{Note7()}]{Note7}%
  \BibitemOpen
  \bibinfo {note} {The apparent peak of $\tau _{_{\protect \mathrm {B}}}$ at
  $T_{\protect \mathrm {c}}$ in the Co-7\% sample is due to the appearance of
  the SC relaxation component.}\BibitemShut {Stop}%
\bibitem [{Note8()}]{Note8}%
  \BibitemOpen
  \bibinfo {note} {At high temperature it is strongly influenced by the
  acoustic shock wave feature around the 10 ps delay.}\BibitemShut {Stop}%
\bibitem [{Note9()}]{Note9}%
  \BibitemOpen
  \bibinfo {note} {The SC component shows the same $T$-dependence for both
  polarizations consistent with isotropic SC gaps.}\BibitemShut {Stop}%
\bibitem [{\citenamefont {Dvorsek}\ \emph {et~al.}(2002)\citenamefont
  {Dvorsek}, \citenamefont {Kabanov}, \citenamefont {Demsar}, \citenamefont
  {Kazakov}, \citenamefont {Karpinski},\ and\ \citenamefont
  {Mihailovic}}]{DvorsekKabanov2002}%
  \BibitemOpen
  \bibfield  {author} {\bibinfo {author} {\bibfnamefont {D.}~\bibnamefont
  {Dvorsek}}, \bibinfo {author} {\bibfnamefont {V.~V.}\ \bibnamefont
  {Kabanov}}, \bibinfo {author} {\bibfnamefont {J.}~\bibnamefont {Demsar}},
  \bibinfo {author} {\bibfnamefont {S.~M.}\ \bibnamefont {Kazakov}}, \bibinfo
  {author} {\bibfnamefont {J.}~\bibnamefont {Karpinski}}, \ and\ \bibinfo
  {author} {\bibfnamefont {D.}~\bibnamefont {Mihailovic}},\ }\href {\doibase
  10.1103/PhysRevB.66.020510} {\bibfield  {journal} {\bibinfo  {journal} {Phys.
  Rev. B}\ }\textbf {\bibinfo {volume} {66}},\ \bibinfo {pages} {020510}
  (\bibinfo {year} {2002})}\BibitemShut {NoStop}%
\bibitem [{\citenamefont {Kabanov}\ \emph {et~al.}(1999)\citenamefont
  {Kabanov}, \citenamefont {Demsar}, \citenamefont {Podobnik},\ and\
  \citenamefont {Mihailovic}}]{KabanovDemsar99}%
  \BibitemOpen
  \bibfield  {author} {\bibinfo {author} {\bibfnamefont {V.~V.}\ \bibnamefont
  {Kabanov}}, \bibinfo {author} {\bibfnamefont {J.}~\bibnamefont {Demsar}},
  \bibinfo {author} {\bibfnamefont {B.}~\bibnamefont {Podobnik}}, \ and\
  \bibinfo {author} {\bibfnamefont {D.}~\bibnamefont {Mihailovic}},\ }\href
  {\doibase 10.1103/PhysRevB.59.1497} {\bibfield  {journal} {\bibinfo
  {journal} {Phys. Rev. B}\ }\textbf {\bibinfo {volume} {59}},\ \bibinfo
  {pages} {1497} (\bibinfo {year} {1999})}\BibitemShut {NoStop}%
\bibitem [{\citenamefont {Analytis}\ \emph {et~al.}(2009)\citenamefont
  {Analytis}, \citenamefont {McDonald}, \citenamefont {Chu}, \citenamefont
  {Riggs}, \citenamefont {Bangura}, \citenamefont {Kucharczyk}, \citenamefont
  {Johannes},\ and\ \citenamefont {Fisher}}]{AnalytisMcDonald2009}%
  \BibitemOpen
  \bibfield  {author} {\bibinfo {author} {\bibfnamefont {J.}~\bibnamefont
  {Analytis}}, \bibinfo {author} {\bibfnamefont {R.}~\bibnamefont {McDonald}},
  \bibinfo {author} {\bibfnamefont {J.}~\bibnamefont {Chu}}, \bibinfo {author}
  {\bibfnamefont {S.}~\bibnamefont {Riggs}}, \bibinfo {author} {\bibfnamefont
  {A.}~\bibnamefont {Bangura}}, \bibinfo {author} {\bibfnamefont
  {C.}~\bibnamefont {Kucharczyk}}, \bibinfo {author} {\bibfnamefont
  {M.}~\bibnamefont {Johannes}}, \ and\ \bibinfo {author} {\bibfnamefont
  {I.}~\bibnamefont {Fisher}},\ }\href@noop {} {\bibfield  {journal} {\bibinfo
  {journal} {Physical Review B}\ }\textbf {\bibinfo {volume} {80}},\ \bibinfo
  {pages} {64507} (\bibinfo {year} {2009})}\BibitemShut {NoStop}%
\bibitem [{Note10()}]{Note10}%
  \BibitemOpen
  \bibinfo {note} {The $T$-dependence of the $\protect \mathcal {P}^{-}$
  $\Delta R/R$ magnitude is qualitatively similar with an offset due to another
  relaxation process.}\BibitemShut {Stop}%
\bibitem [{\citenamefont {Hu}\ \emph {et~al.}(2008)\citenamefont {Hu},
  \citenamefont {Dong}, \citenamefont {Li}, \citenamefont {Li}, \citenamefont
  {Zheng}, \citenamefont {Chen}, \citenamefont {Luo},\ and\ \citenamefont
  {Wang}}]{HuDong2008}%
  \BibitemOpen
  \bibfield  {author} {\bibinfo {author} {\bibfnamefont {W.}~\bibnamefont
  {Hu}}, \bibinfo {author} {\bibfnamefont {J.}~\bibnamefont {Dong}}, \bibinfo
  {author} {\bibfnamefont {G.}~\bibnamefont {Li}}, \bibinfo {author}
  {\bibfnamefont {Z.}~\bibnamefont {Li}}, \bibinfo {author} {\bibfnamefont
  {P.}~\bibnamefont {Zheng}}, \bibinfo {author} {\bibfnamefont
  {G.}~\bibnamefont {Chen}}, \bibinfo {author} {\bibfnamefont {J.}~\bibnamefont
  {Luo}}, \ and\ \bibinfo {author} {\bibfnamefont {N.}~\bibnamefont {Wang}},\
  }\href@noop {} {\bibfield  {journal} {\bibinfo  {journal} {Phys Rev Lett}\
  }\textbf {\bibinfo {volume} {101}},\ \bibinfo {pages} {257005} (\bibinfo
  {year} {2008})}\BibitemShut {NoStop}%
\bibitem [{\citenamefont {Lucarelli}\ \emph {et~al.}(2010)\citenamefont
  {Lucarelli}, \citenamefont {Dusza}, \citenamefont {Pfuner}, \citenamefont
  {Lerch}, \citenamefont {Analytis}, \citenamefont {Chu}, \citenamefont
  {Fisher},\ and\ \citenamefont {Degiorgi}}]{LucarelliDusza2010}%
  \BibitemOpen
  \bibfield  {author} {\bibinfo {author} {\bibfnamefont {A.}~\bibnamefont
  {Lucarelli}}, \bibinfo {author} {\bibfnamefont {A.}~\bibnamefont {Dusza}},
  \bibinfo {author} {\bibfnamefont {F.}~\bibnamefont {Pfuner}}, \bibinfo
  {author} {\bibfnamefont {P.}~\bibnamefont {Lerch}}, \bibinfo {author}
  {\bibfnamefont {J.~G.}\ \bibnamefont {Analytis}}, \bibinfo {author}
  {\bibfnamefont {J.-H.}\ \bibnamefont {Chu}}, \bibinfo {author} {\bibfnamefont
  {I.~R.}\ \bibnamefont {Fisher}}, \ and\ \bibinfo {author} {\bibfnamefont
  {L.}~\bibnamefont {Degiorgi}},\ }\href
  {http://stacks.iop.org/1367-2630/12/i=7/a=073036} {\bibfield  {journal}
  {\bibinfo  {journal} {New Journal of Physics}\ }\textbf {\bibinfo {volume}
  {12}},\ \bibinfo {pages} {073036} (\bibinfo {year} {2010})}\BibitemShut
  {NoStop}%
\bibitem [{\citenamefont {Kabanov}\ and\ \citenamefont
  {Alexandrov}(2008)}]{KabanovAlexandrov2008}%
  \BibitemOpen
  \bibfield  {author} {\bibinfo {author} {\bibfnamefont {V.~V.}\ \bibnamefont
  {Kabanov}}\ and\ \bibinfo {author} {\bibfnamefont {A.~S.}\ \bibnamefont
  {Alexandrov}},\ }\href {\doibase 10.1103/PhysRevB.78.174514} {\bibfield
  {journal} {\bibinfo  {journal} {Physical Review B (Condensed Matter and
  Materials Physics)}\ }\textbf {\bibinfo {volume} {78}},\ \bibinfo {eid}
  {174514} (\bibinfo {year} {2008})}\BibitemShut {NoStop}%
\bibitem [{\citenamefont {Gadermaier}\ \emph {et~al.}(2010)\citenamefont
  {Gadermaier}, \citenamefont {Alexandrov}, \citenamefont {Kabanov},
  \citenamefont {Kusar}, \citenamefont {Mertelj}, \citenamefont {Yao},
  \citenamefont {Manzoni}, \citenamefont {Brida}, \citenamefont {Cerullo},\
  and\ \citenamefont {Mihailovic}}]{GadermaierAlexandrov2010}%
  \BibitemOpen
  \bibfield  {author} {\bibinfo {author} {\bibfnamefont {C.}~\bibnamefont
  {Gadermaier}}, \bibinfo {author} {\bibfnamefont {A.~S.}\ \bibnamefont
  {Alexandrov}}, \bibinfo {author} {\bibfnamefont {V.~V.}\ \bibnamefont
  {Kabanov}}, \bibinfo {author} {\bibfnamefont {P.}~\bibnamefont {Kusar}},
  \bibinfo {author} {\bibfnamefont {T.}~\bibnamefont {Mertelj}}, \bibinfo
  {author} {\bibfnamefont {X.}~\bibnamefont {Yao}}, \bibinfo {author}
  {\bibfnamefont {C.}~\bibnamefont {Manzoni}}, \bibinfo {author} {\bibfnamefont
  {D.}~\bibnamefont {Brida}}, \bibinfo {author} {\bibfnamefont
  {G.}~\bibnamefont {Cerullo}}, \ and\ \bibinfo {author} {\bibfnamefont
  {D.}~\bibnamefont {Mihailovic}},\ }\href {\doibase
  10.1103/PhysRevLett.105.257001} {\bibfield  {journal} {\bibinfo  {journal}
  {Phys. Rev. Lett.}\ }\textbf {\bibinfo {volume} {105}},\ \bibinfo {pages}
  {257001} (\bibinfo {year} {2010})}\BibitemShut {NoStop}%
\bibitem [{\citenamefont {Mittal}\ \emph {et~al.}(2008)\citenamefont {Mittal},
  \citenamefont {Su}, \citenamefont {Rols}, \citenamefont {Chatterji},
  \citenamefont {Chaplot}, \citenamefont {Schober}, \citenamefont {Rotter},
  \citenamefont {Johrendt},\ and\ \citenamefont {Brueckel}}]{MittalSu2008}%
  \BibitemOpen
  \bibfield  {author} {\bibinfo {author} {\bibfnamefont {R.}~\bibnamefont
  {Mittal}}, \bibinfo {author} {\bibfnamefont {Y.}~\bibnamefont {Su}}, \bibinfo
  {author} {\bibfnamefont {S.}~\bibnamefont {Rols}}, \bibinfo {author}
  {\bibfnamefont {T.}~\bibnamefont {Chatterji}}, \bibinfo {author}
  {\bibfnamefont {S.~L.}\ \bibnamefont {Chaplot}}, \bibinfo {author}
  {\bibfnamefont {H.}~\bibnamefont {Schober}}, \bibinfo {author} {\bibfnamefont
  {M.}~\bibnamefont {Rotter}}, \bibinfo {author} {\bibfnamefont
  {D.}~\bibnamefont {Johrendt}}, \ and\ \bibinfo {author} {\bibfnamefont
  {T.}~\bibnamefont {Brueckel}},\ }\href {\doibase 10.1103/PhysRevB.78.104514}
  {\bibfield  {journal} {\bibinfo  {journal} {Phys. Rev. B}\ }\textbf {\bibinfo
  {volume} {78}},\ \bibinfo {pages} {104514} (\bibinfo {year}
  {2008})}\BibitemShut {NoStop}%
\bibitem [{\citenamefont {Mihailovic}\ and\ \citenamefont
  {Demsar}(1999)}]{MihailovicDemsar99}%
  \BibitemOpen
  \bibfield  {author} {\bibinfo {author} {\bibfnamefont {D.}~\bibnamefont
  {Mihailovic}}\ and\ \bibinfo {author} {\bibfnamefont {J.}~\bibnamefont
  {Demsar}},\ }\enquote {\bibinfo {title} {{Spectroscopy of Superconducting
  Materials}},}\ \ (\bibinfo  {publisher} {American Chemical Society:
  Washington, DC},\ \bibinfo {year} {1999})\ Chap.\ \bibinfo {chapter}
  {{Time-resolved optical studies of quasiparticle dynamics in high-temperature
  superconductors}}, pp.\ \bibinfo {pages} {230--244}\BibitemShut {NoStop}%
\bibitem [{Note11()}]{Note11}%
  \BibitemOpen
  \bibinfo {note} {The Co-2.5\% sample shows a clear two component initial
  decay at the room temperature so a two exponential fit was used.}\BibitemShut
  {Stop}%
\bibitem [{\citenamefont {Tu}\ \emph {et~al.}(2010)\citenamefont {Tu},
  \citenamefont {Li}, \citenamefont {Liu}, \citenamefont {Punnoose},
  \citenamefont {Gong}, \citenamefont {Ren}, \citenamefont {Li}, \citenamefont
  {Cao}, \citenamefont {Xu},\ and\ \citenamefont {Homes}}]{TuLi2010}%
  \BibitemOpen
  \bibfield  {author} {\bibinfo {author} {\bibfnamefont {J.~J.}\ \bibnamefont
  {Tu}}, \bibinfo {author} {\bibfnamefont {J.}~\bibnamefont {Li}}, \bibinfo
  {author} {\bibfnamefont {W.}~\bibnamefont {Liu}}, \bibinfo {author}
  {\bibfnamefont {A.}~\bibnamefont {Punnoose}}, \bibinfo {author}
  {\bibfnamefont {Y.}~\bibnamefont {Gong}}, \bibinfo {author} {\bibfnamefont
  {Y.~H.}\ \bibnamefont {Ren}}, \bibinfo {author} {\bibfnamefont {L.~J.}\
  \bibnamefont {Li}}, \bibinfo {author} {\bibfnamefont {G.~H.}\ \bibnamefont
  {Cao}}, \bibinfo {author} {\bibfnamefont {Z.~A.}\ \bibnamefont {Xu}}, \ and\
  \bibinfo {author} {\bibfnamefont {C.~C.}\ \bibnamefont {Homes}},\ }\href
  {\doibase 10.1103/PhysRevB.82.174509} {\bibfield  {journal} {\bibinfo
  {journal} {Phys. Rev. B}\ }\textbf {\bibinfo {volume} {82}},\ \bibinfo
  {pages} {174509} (\bibinfo {year} {2010})}\BibitemShut {NoStop}%
\bibitem [{\citenamefont {Nakajima}\ \emph {et~al.}(2010)\citenamefont
  {Nakajima}, \citenamefont {Ishida}, \citenamefont {Kihou}, \citenamefont
  {Tomioka}, \citenamefont {Ito}, \citenamefont {Yoshida}, \citenamefont {Lee},
  \citenamefont {Kito}, \citenamefont {Iyo}, \citenamefont {Eisaki},
  \citenamefont {Kojima},\ and\ \citenamefont {Uchida}}]{NakajimaIshida2010}%
  \BibitemOpen
  \bibfield  {author} {\bibinfo {author} {\bibfnamefont {M.}~\bibnamefont
  {Nakajima}}, \bibinfo {author} {\bibfnamefont {S.}~\bibnamefont {Ishida}},
  \bibinfo {author} {\bibfnamefont {K.}~\bibnamefont {Kihou}}, \bibinfo
  {author} {\bibfnamefont {Y.}~\bibnamefont {Tomioka}}, \bibinfo {author}
  {\bibfnamefont {T.}~\bibnamefont {Ito}}, \bibinfo {author} {\bibfnamefont
  {Y.}~\bibnamefont {Yoshida}}, \bibinfo {author} {\bibfnamefont {C.~H.}\
  \bibnamefont {Lee}}, \bibinfo {author} {\bibfnamefont {H.}~\bibnamefont
  {Kito}}, \bibinfo {author} {\bibfnamefont {A.}~\bibnamefont {Iyo}}, \bibinfo
  {author} {\bibfnamefont {H.}~\bibnamefont {Eisaki}}, \bibinfo {author}
  {\bibfnamefont {K.~M.}\ \bibnamefont {Kojima}}, \ and\ \bibinfo {author}
  {\bibfnamefont {S.}~\bibnamefont {Uchida}},\ }\href {\doibase
  10.1103/PhysRevB.81.104528} {\bibfield  {journal} {\bibinfo  {journal} {Phys.
  Rev. B}\ }\textbf {\bibinfo {volume} {81}},\ \bibinfo {pages} {104528}
  (\bibinfo {year} {2010})}\BibitemShut {NoStop}%
\bibitem [{Note12()}]{Note12}%
  \BibitemOpen
  \bibinfo {note} {Our samples have the same origin as in Ref. {[}\protect
  \rev@citealpnum {LucarelliDusza2010}{]} and $\protect \nicefrac {1}{\tau
  _{_{\protect \mathrm {IB}}}}$ can be strongly affected by
  impurities.}\BibitemShut {Stop}%
\bibitem [{Note13()}]{Note13}%
  \BibitemOpen
  \bibinfo {note} {C. Gadermaier \protect \emph {et al.}, unpublished
  data.}\BibitemShut {Stop}%
\end{thebibliography}%

\end{document}